\documentclass{aa}  

\usepackage{graphicx}
\usepackage{array}
\usepackage{txfonts}
\usepackage{caption}
\usepackage{hyperref}
\usepackage{url}
\usepackage{natbib}
\usepackage{breqn}
\usepackage{booktabs}
\usepackage{multirow}
\usepackage{siunitx}
\bibpunct{(}{)}{;}{a}{}{,} 
\usepackage[Table,usenames,dvipsnames]{xcolor}
\newcolumntype{C}[1]{>{\centering\let\newline\\\arraybackslash\hspace{0pt}}m{#1}}
\begin{document} 
\title{Synthetic observations of spiral arm tracers\\ of a simulated Milky Way analog}
\author{Stefan Reissl\inst{\ref{inst1}}, Jeroen M. Stil\inst{\ref{inst2}}, En Chen \inst{\ref{inst3}}, Robin G. Treß\inst{\ref{inst1}}, Mattia C. Sormani\inst{\ref{inst1}}, Rowan J. Smith \inst{\ref{inst3}},  Ralf S. Klessen\inst{\ref{inst1},\ref{inst4}}, \\ Megan Buick\inst{\ref{inst2}}, Simon C. O. Glover\inst{\ref{inst1}}, Russell Shanahan\inst{\ref{inst2}}, Stephen J. Lemmer\inst{\ref{inst2}}, Juan D. Soler \inst{\ref{inst5}}, \\Henrik Beuther \inst{\ref{inst5}}, James S. Urquhart \inst{\ref{inst6}} , L.D. Anderson \inst{\ref{inst7},\ref{inst8},\ref{inst9}}, Karl M. Menten \inst{\ref{inst10}},\\ Andreas Brunthaler \inst{\ref{inst10}}, Sarah Ragan  \inst{\ref{inst11}}, \and Michael R. Rugel \inst{\ref{inst10}}}
\institute{
\centering \label{inst1} Universit{\"a}t Heidelberg, Zentrum f{\"u}r Astrononmie, Institut f{\"u}r Theoretische Astrophysik, Albert-Ueberle-Str. 2, 69120 Heidelberg, Germany
\and
\centering \label{inst2} Department of Physics and Astronomy, The University of Calgary, 2500 University Drive NW, Calgary AB T2N 1N4, Canada
\and
\centering \label{inst3} Jodrell Bank Centre for Astrophysics, School of Physics and  Astronomy, University of Manchester, Oxford Road, Manchester, M13 9PL,UK
\and
\centering \label{inst4} Universit{\"a}t Heidelberg, Interdisziplin{\"a}res Zentrum f{\"u}r Wissenschaftliches Rechnen, Im Neuenheimer Feld 205, 69120 Heidelberg, Germany
\and
\centering \label{inst5} Max Planck Institute for Astronomy, Königstuhl 17, D-69117 Heidelberg, Germany
\and
\centering \label{inst6} Centre for Astrophysics and Planetary Science, University of Kent, Canterbury CT2 7NH, UK
\and
\centering \label{inst7} Department of Physics and Astronomy, West Virginia University, Morgantown, WV 26506, USA
\and
\centering \label{inst8} Adjunct Astronomer at the Green Bank Observatory, P.O. Box 2, Green Bank, WV 24944, USA
\and
\centering \label{inst9} Center for Gravitational Waves and Cosmology, West Virginia University, Chestnut Ridge Research Building, Morgantown, WV 26505, USA
\and
\centering \label{inst10} Max-Planck-Institut für Radioastronomie, Auf dem Hügel 69, D-53121 Bonn, German
\and
\centering \label{inst11} School of Physics \& Astronomy, Cardiff University, The Parade, Cardiff CF24 3AA, UK
}

\abstract
   {{ The Faraday rotation measure ($RM$) is often used to study the magnetic field strength and orientation within the ionized medium of the Milky Way.} Recent observations indicate an $RM$ magnitude in the spiral arms that exceeds the commonly assumed range. This raises the question of how and under what conditions spiral arms create such strong Faraday rotation.}
{We investigate the effect of spiral arms on Galactic Faraday rotation through shock compression of the interstellar medium. { It has recently been suggested that the Sagittarius spiral arm creates a strong peak in Faraday rotation where the line of sight is tangent to the arm, and that enhanced Faraday rotation follows along side lines which intersect the arm. Here, we seek to understand the physical conditions that may give rise to this effect and the role of viewing geometry.}}
   {{ We apply a magnetohydrodynamic simulation of the multi-phase interstellar medium in a Milky Way type spiral galaxy disk in combination with radiative transfer in order to evaluate different tracers of spiral arm structures.} For observers embedded in the disk, dust intensity, synchrotron emission and the kinematics of molecular gas observations are derived to identify which spiral arm tangents are observable. Faraday rotation measures are calculated through the disk and evaluated in the context of different observer positions.  { The observer's perspectives are related to the parameters of the local bubble surrounding the observer and their contribution to the total Faraday rotation measure along the line of sight.}}
   {We reproduce a scattering of tangent points for the different tracers of about $6^\circ$ per spiral arm similar to the Milky Way. As for the $RM$, the model shows that compression of the interstellar medium and associated amplification of the magnetic field in spiral arms enhances Faraday rotation by a few hundred rad m$^{-2}$ on top of the mean contribution of the disk. The arm-inter-arm contrast in Faraday rotation per unit distance along the line of sight is approximately $\sim 10$ in the inner Galaxy, fading to $\sim 2$ in the outer Galaxy in tandem with the waning contrast of other tracers of spiral arms. We identify a shark-fin like pattern in the $RM$ Milky Way observations as well as the synthetic data that is characteristic for a galaxy with spiral arms.}
 {}
  \titlerunning{Simulations of Spiral Arm Tracers in the Milky Way}
  \authorrunning{Reissl et al.}
  \maketitle
%

\section{Introduction}
The magnetic field of the Milky Way affects the physics of the interstellar medium on various scales \citep{Klessen2016}. In turn, the magnetic field is shaped and amplified by motions in the interstellar medium \citep{Beck2015}, including turbulent motions injected by stellar winds and supernova explosions \citep{Ferriere1991,Tomisaka1998}, and streaming motions of gas. In particular, spiral arms can shape the magnetic field on a galactic scale as the gas entering the arms is compressed in a shock \citep{Roberts1969}. The strength and the structure of the magnetic field in spiral arms are important initial conditions for theories of star formation, cosmic ray diffusion, and dynamics of the interstellar medium, for example the formation of filaments \citep{Inoue2018} and the Parker instability \citep{Koertgen2018}.  

{ The detection of spiral arms within the Milky Way is aggravated by the fact that the view towards the Milky Way's spiral arms and the Galactic center is naturally blocked. Due to the position of our own solar system within the Galactic disk dust extinction at optical and UV wavelengths are hampered.} However, observations of synchrotron radiation and far-infrared dust emission \citep{Beuermann1985, Beuther2012}, VLBI parallaxes of maser sources associated with high-mass star forming regions \citep{Reid2019}  as well as Faraday rotation measurements \citep{Shanahan2019} open a window to infer the spiral structure of our home galaxy. In detail, the observations of diffuse synchrotron emission in well-resolved face-on galaxies show evidence for amplified ordered magnetic fields associated with tracers of spiral arm shocks \citep[e.g.][]{Jansson2012}. The polarized intensity shows ordering of the magnetic field in the plane of the sky, in a direction perpendicular to the plane of polarization. This fact cannot be used to prove the presence of large-scale magnetic fields, because the polarized signal of a compressed turbulent magnetic field is the same as that of a turbulent field superposed on a large-scale magnetic field \citep{Sokoloff1998}. Faraday rotation in turn can reveal large-scale magnetic fields \citep{Berkhuijsen2003,Gaensler2005} and magnetic field reversals \citep[e.g.][]{SimardNormandin1979,Mora-Partiarroyo2019}.

The large-scale magnetic field of the Milky Way can be studied in much greater detail with all-sky dust polarimetry 
\citep[e.g.][]{PlanckXIX2015,Planck2018XII}. Further details about the Milky Way's magnetic field can be gathered by Faraday rotation observations of pulsars and polarized extragalactic radio sources \citep{Han2006,Jansson2012}. Such Faraday rotation observations reveal that the magnetic field in the Sagittarius arm is reversed with respect to the main magnetic field direction in the inner Galaxy. The existence of other magnetic field reversals is the subject of continued debate. 

Recently, \citet{Shanahan2019} presented a pattern of Faraday rotation along the Galactic plane based on data provided by the THOR survey \citep[see][for details]{Beuther2016,Wang2020}. Here, \citet{Shanahan2019} suggests the Sagittarius arm is a much stronger Faraday screen within the Milky Way than previously thought. This pattern consists of a strong peak in Faraday rotation associated with the spiral arm tangent, flanked by enhanced but lower Faraday rotation along lines of sight through the inner Galaxy that intersect the arm, and much less Faraday rotation along lines of sight beyond the tangent, that do not intersect the arm.

Further insight into the magnetic field structure of spiral arms can be obtained if the observed pattern can be reproduced by numerical models of the flow of magnetized interstellar medium through the spiral arms \citep{Fletcher2011}. Such simulations can provide insight into the importance of viewing geometry, while connecting key physics to observable quantities without the need to deproject the observed quantities. This requires implementation of emission and radiation transfer based on local conditions defined by the simulation. Direct comparison of MHD simulations with Faraday rotation data on a Galactic scale are exceedingly rare, while a some studies have addressed MHD
turbulence without Galactic structure, e.g., of super bubbles \citep[][]{Pakmor2018}.

What is required to relate the simulations to observations of Galactic Faraday rotation? The main observed properties are the magnitude and sign of Faraday rotation in relation to other tracers of spiral arms from the perspective of an observer embedded in the disk of the Milky Way. Simulations must therefore reproduce the three-dimensional structure of the multi-phase interstellar medium with a magnetic field, and generate mock observations of key tracers such as dust and synchrotron emission of the multi-phase interstellar medium that can be compared with the Milky Way in a statistical sense. In the Milky Way, spiral arms are best traced by kinematics of the molecular and atomic gas, while directions of spiral arm tangents are traced also by excess emission in the form of radio continuum emission from synchrotron radiation and free-free emission from star formation regions \citep[e.g.][]{Beuermann1985}, and dust emission \citep[e.g.][]{Beuther2012}. Self-consistent treatment of these observables is key to effectively relating simulations to observations.

The extent to which the signatures of large-scale Galactic structure stand out among other effects, such as the structure of the local interstellar medium, is another question that may be addressed with simulations. Mock observations generated by radiative transfer post-processing codes can be evaluated from positions in the simulations that best represent the perspective of an observer on Earth. In this paper, we bridge the gap between simulations of gas dynamics and magnetic field evolution on a Galactic scale with observations of Milky Way like spiral structures through the implementation of emission processes and radiation transfer through ray-tracing. However, we emphasize that key-aspects of such a complex system as the Milky Way are still a field of ongoing research. Naturally, any Milky Way analog may only cover certain physical processes and remains merely an approximation. The Resulting synthetic observations can rarely directly be compared with the original and have to be taken with care.

This paper is structured as follows: in Sect. \ref{sect:Milky WayModel} we summarize our numerical model of the Galactic disk with spiral arms and the physical effects that it includes. In Sect. \ref{sect:RTPostProcessing} we outline the radiative transfer post-processing techniques and the basic physical properties of the different tracers for which synthetic date is generated in this paper. We present the synthetic observations in Sect. \ref{sect:SpiralDetection} and analyze their longitudinal profiles to determine the tangent points of the spiral arms. We temporarily change the perspective to that of an outside observer in Sect. \ref{sect:OutsideObserver} in order to evaluate the validity of our spiral arm detections. In Sect. \ref{sect:OrionSpur} we go back to the perspective of an observer within the Galactic disk and explore the possibility of deriving the pitch angle of our Local Arm in the Milky Way by means of Faraday $RM$ observations. In Sect. \ref{sect:SpiralArmProfile} we offer a purely geometric explanation for the characteristic shark-fin peak profile in the longitudinal profiles. This profile constitutes the observational signature of Faraday rotation at a spiral arm tangent point. In Sect. \ref{sect:OriginRM} we investigate where Faraday rotation originates along the line of sight by tracing the propagation of polarized radio waves from the outside of the model Galaxy toward the observer. In Sect. \ref{sect:LocalBubble}, we investigate the implications that the existence of a Local Bubble has for the detection of the spiral arms. In Sect. \ref{sect:caveats}, we discuss the impact that the simplifications made in our numerical model may have on our results. Finally, we summarize our results in Sect. \ref{sect:Summary}.

\section{The Milky Way model}
\label{sect:Milky WayModel}
{ In this section we outline the numerical methods and the physical principles for constructing our Milky Way-disk analog.}

\subsection{Numerical Methods}
\label{sect:NumMethods}
We perform a Galactic-scale MHD simulation using the moving-mesh code \textsc{Arepo} \citep{Springel2010} to model the magnetised interstellar medium (ISM). The simulations include gas self-gravity, magnetic fields, a live chemical network, sink particles to represent star formation, supernovae (SNe) feedback and an externally imposed four-armed spiral potential.

{ This galaxy model forms part of the ‘Cloud Factory’ simulation suite recently presented in \cite{Smith2020}, which aims to investigate the connection between galactic-scale forces and star formation in individual molecular clouds. We employ an idealized externally imposed galactic potential that consists of an axisymmetric part plus a spiral perturbation. The axisymmetric part is composed of the sum of a stellar disc, bulge and dark matter halo according to the best fitting model of \cite{Mcmillan2017}. A spiral perturbation as introduced by \cite{Cox2002} is added to this potential to give four well-defined spiral arms resembling the Milky Way (compare Fig. 1, and \citealt{Reid2019}). The spiral potential has a pitch angle\footnote{We note that the pitch angle is sometimes defined in literature with respect to the magnetic field direction. Here, the pitch angle is the deviation of the spiral arm with respect to a perfect circle around the Galactic center i.e. $\Psi = 0^\circ$.} $\Psi=15^{\circ}$, and a pattern speed of $2\times 10^{-8}\ \mathrm{rad\ yr}^{-1}$, which corresponds to a co-rotation radius of $11\ \mathrm{kpc}$ (see also \citealt{Smith2014a}).}

The ISM is assumed to be composed of atomic hydrogen and helium, together with carbon and oxygen, with abundances matching those in the local ISM. The chemical evolution of the gas is followed using the NL97 chemical network of \citet{Glover2012}, which has been extensively used in previous numerical studies of the ISM \citep[see e.g.][]{Glover2012,Smith2014a,Walch2015,Sormani2018,Tress2019,Smith2020}. This network combines the treatment of hydrogen chemistry introduced in \citet{Glover2007} with a simplified model  for CO formation and destruction taken from \citet{Nelson1997}. The NL97 network includes H$_{2}$ formation on grains, H$_{2}$ photodissociation, collisional dissociation, H$^{+}$ recombination on grains and in the gas phase (see Table 1 of \citealt{Glover2007}), a highly approximate treatment of CO formation, CO photodissociation, and cosmic ray ionisation of H and H$_2$. The only charged species traced in the NL97 network are H$^{+}$ and C$^{+}$, and so the density of free electrons is given by $n_{\rm el} = n_{\rm H^{+}} + n_{\rm C^{+}}$. A full description of the NL97 network can be found in \citet{Glover2012}.

For simplicity, we assume a uniform interstellar radiation field (ISRF), with a intensity and spectrum given by \citet{Draine1978}. The attenuation of this radiation field in dense molecular clouds owing to dust shielding and H$_{2}$ self-shielding is modelled using the TreeCol algorithm of \citet{Clark2012}. The cosmic ray ionisation rate is also assumed to be spatially constant, with a value of $\zeta_{\rm H} = 3\times 10^{-17} \: {\rm s^{-1}}$ for atomic hydrogen . Ionisation rates for other chemical species are scaled from this value as described in \citet{Glover2012}.

Star formation is represented by non-gaseous sink particles, which are formed from gravitationally bound gas above a threshold gas density of $n_{\mathrm{g}} = 10^3\ \mathrm{cm}^{-3}$. The sinks can accrete gas and form stars with a star formation efficiency of $2\%$, which is consistent with observations in GMCs in the Milky Way \citep[e.g.][]{Krumholz2007}. The sinks represent small clusters of stars (the maximum stellar content of the sinks is set to $200\ \mathrm{M}_{\odot}$ in this run) and have SN feedback tied to them as outlined in \citet{Tress2019}. Here, we also include a random SN component in the simulations to represent Type Ia supernovae. Full details of the numerical methods we use can be found in \citet{Smith2020}. However, \cite{Smith2020} present only hydrodynamic models. The models used in this paper include ideal MHD using the setup applied in the follow-up paper by \citet{Chen2020}. To do this we use the MHD capabilities of \textsc{Arepo}, as described in \citet{Pakmor2013}, which uses the Powell divergence control scheme \citep{Powell1999}. This methodology has previously been successfully used to model galactic-scale magnetic fields in \citet{Pakmor2017}.

\begin{figure}
\centering
\begin{minipage}[c]{1.0\linewidth}
     \includegraphics[width=1.0\textwidth]{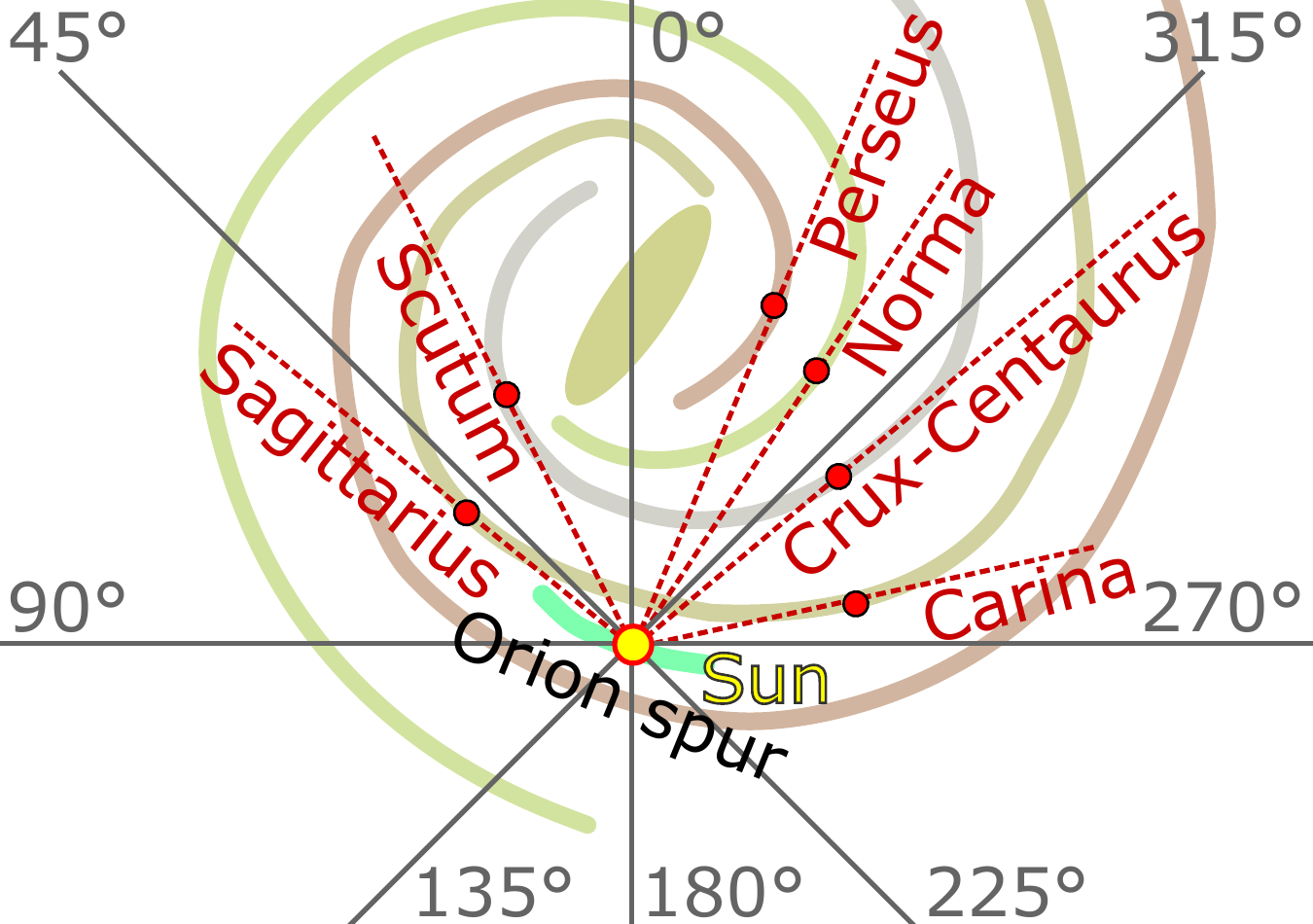}
\end{minipage}
\caption{Sketch of the Milky Way's spiral structure. The spiral arms are indicated in separate colors. Our solar system is placed in the Orion spur about $8.2\ \mathrm{kpc}$ away from the Galactic center. Together with the Milky Way's center the Sun defines the Galactic coordinate system where angles are counted counter clockwise along the longitude.}
\label{fig:Sketch}
\end{figure}

\begin{figure*}
\begin{center}

     \includegraphics[width=0.49\textwidth]{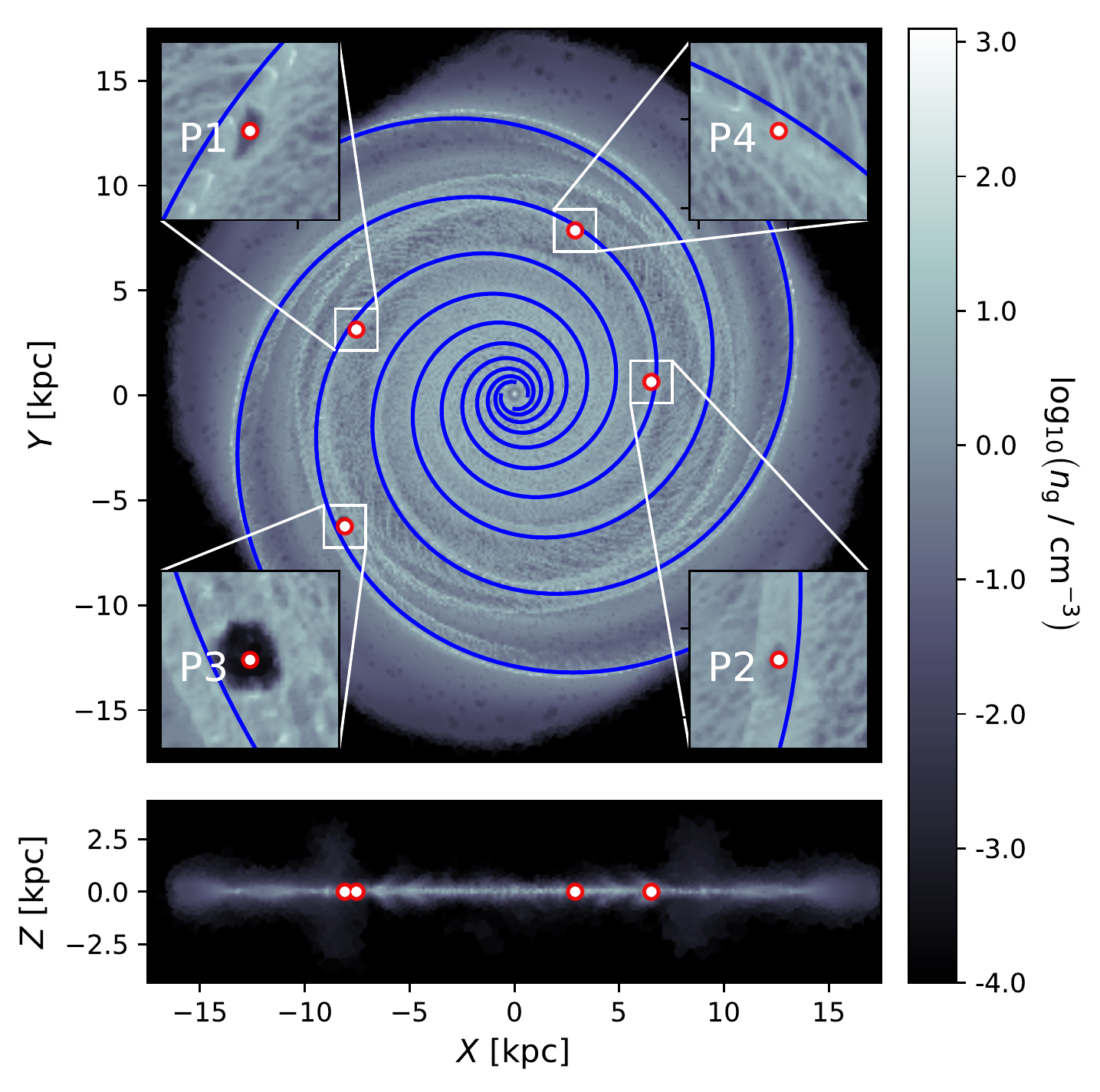}
     \includegraphics[width=0.49\textwidth]{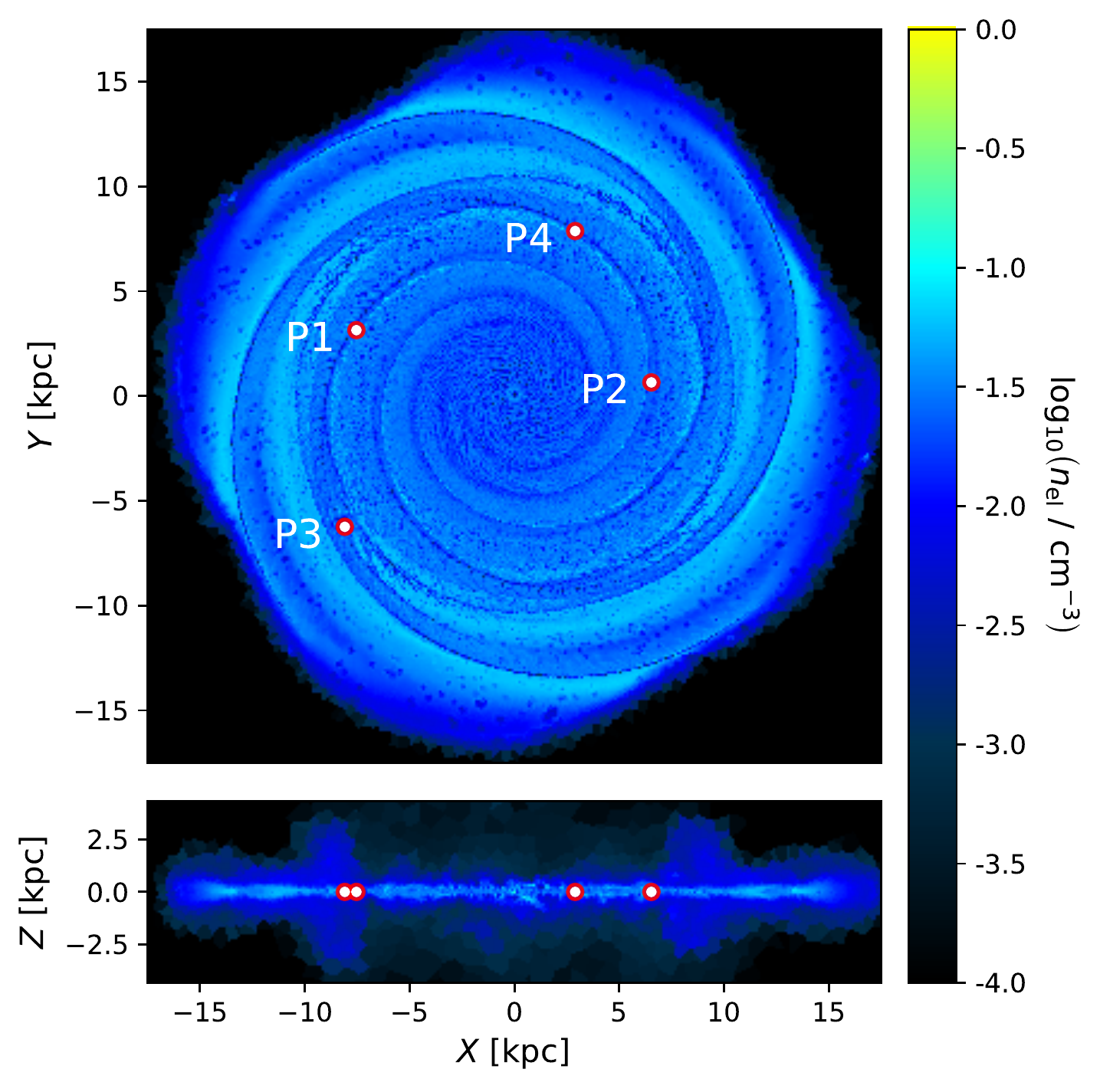}\\

     \includegraphics[width=0.49\textwidth]{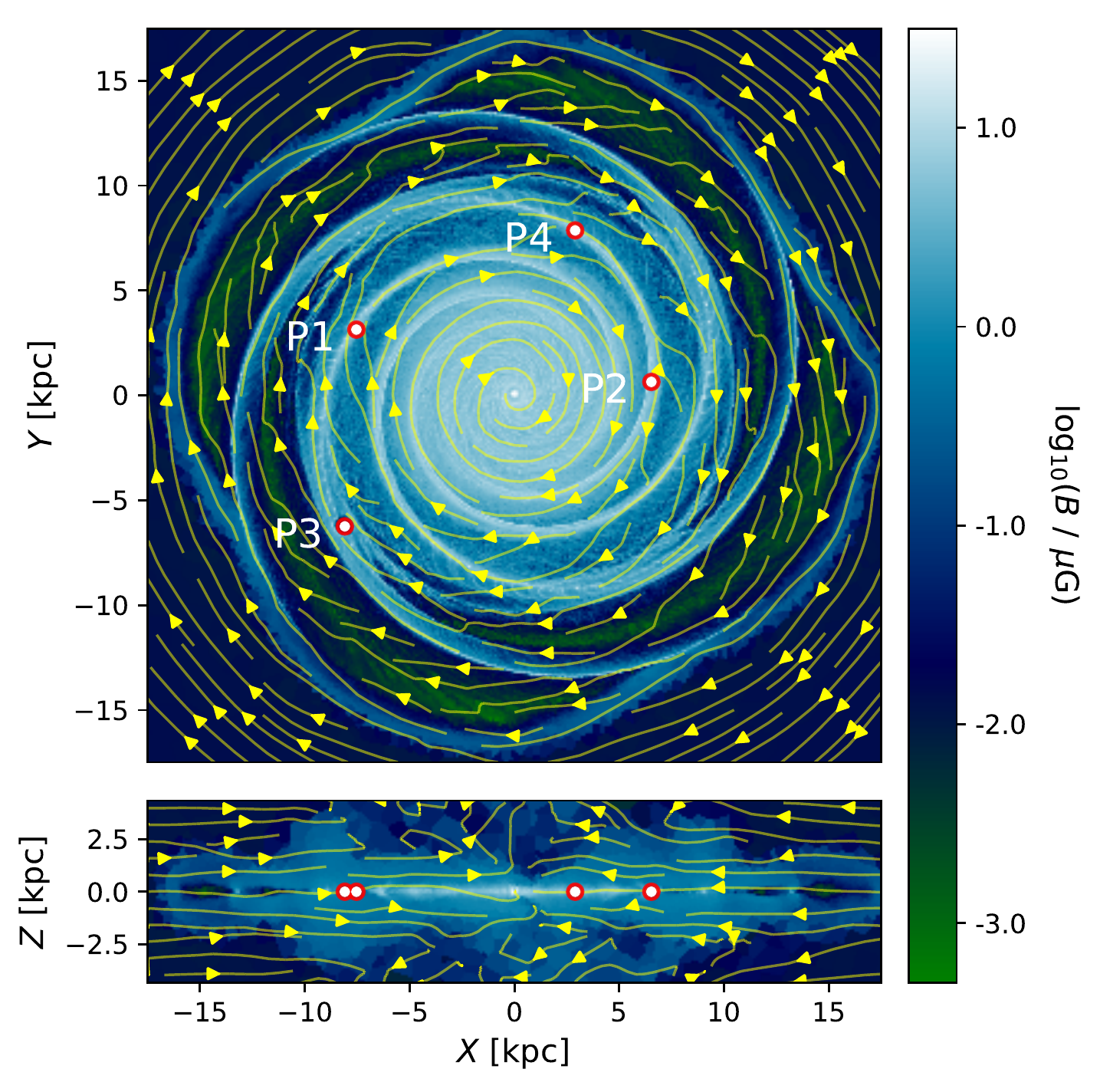}
     \includegraphics[width=0.49\textwidth]{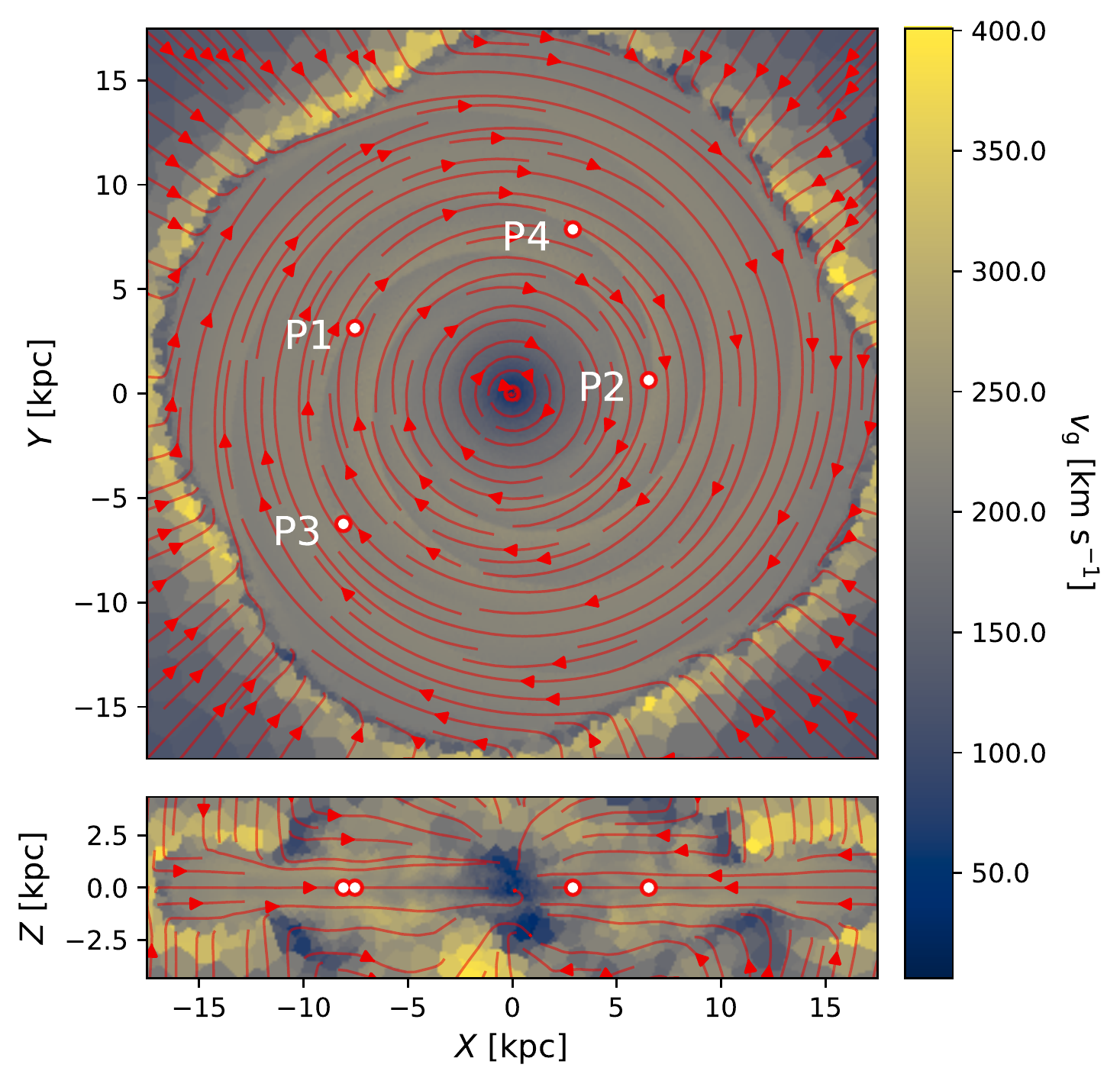}

\end{center}

\caption{ {  Slices though the Milky Way model described in Sect. \ref{sect:SimSetup}. From top-left to bottom-right in clockwise direction: gas density $n_{\mathrm{g}}$, electron density $n_{\mathrm{el}}$, velocity $v_{\mathrm{g}}$, and magnetic field strength $B$. The red circles and white labels indicate the observer positions considered in the paper. The zoom-in panels in the $n_{\mathrm{g}}$ map (top-left) show the local environment for each of the observers P1, P2, P3, and P4 respectively, and the blue spirals indicate the potential of our model. The yellow and red vector fields show the orientation of the magnetic field (bottom-left) and the velocity field (bottom-right), respectively.}}
\label{fig:Midplanes}
\end{figure*}

\subsection{Simulation Setup}
\label{sect:SimSetup}
The initial setup is generated using the following procedure. We assign disk gas densities according to the \textsc{Hi} thin and thick disk components of \cite{Mcmillan2017} to be consistent with our analytic potential. The disk is truncated at a radius of $15\ \mathrm{kpc}$ and surrounded by a hot diffuse medium for reasons of computational efficiency. We allocate the initial velocity for every gas cell following the rotation curve that arises from our analytic potential ($0\ \mathrm{km\ s}^{-1}$ at the galactic center, and $220\ \mathrm{km\ s}^{-1}$ from $4\ \mathrm{kpc}$ outwards).

The run has two phases. Initially for the first $200\ \mathrm{Myr}$, we switch off gas self-gravity, and the sink particles and SN feedback modules but include the galactic potential. Doing this allows the galaxy to become relaxed, the field to undergo an initial amplification, and the spiral arms to be generated. The strength of the seed field is set to the commonly assumed value of $0.04\ \mu\mathrm{G}$ \citep[e.g.][]{Pakmor2017} in a toroidal fashion. During this period, we set the resolution refinement criteria for the Voronoi cells to a target mass of $10^4\ \mathrm{M}_{\odot}$. 

After $200\ \mathrm{Myr}$, when the simulations reach a quasi-steady state, we turn on self-gravity, together with the sink particles and SN feedback modules to allow the gas to collapse to form stars and then be disrupted by feedback. We run with this physics for a further $50\ \mathrm{Myr}$ and use a higher resolution target mass with a maximum of $10^3\ \mathrm{M}_{\odot}$. We also require that the Jeans length \citep{Jeans1902} is resolved by a minimum of four Voronoi cells necessary to avoid artificial fragmentation \citep[][]{Truelove1997}. In this way we reach much higher mass resolution at GMC scales.
The typical spatial resolution in this phase varies with the gas density, but for gas at $n_{\mathrm{g}}\sim 100\ \mathrm{cm}^{-3}$ corresponds to a cell size of about $L\approx 1\ \mathrm{pc}$. All our following analysis is performed using data from a snapshot produced after a total simulation time of $250\ \mathrm{Myr}$, i.e. $50\ \mathrm{Myr}$ after we turn on self-gravity.

Observational tracers such as dust and synchrotron emission are highly sensitive to the local conditions \citep[][]{Reissl2019}. For the purpose of the follow up radiative transfer post-processing we manually added three additional bubbles driven by clustered supernovae chosen to be roughly at Solar distances from the Galactic centre, each associated to a spiral arm. At these locations at $t=225\ \mathrm{Myr}$ the number of SNe associated to a local sink particle is increased to 500. These then explode with a roughly even distribution in time within the following $10\ \mathrm{Myr}$, leading to density cavities representing different stages of bubble evolution. We place three observers at positions within these SN bubbles. Additionally, we hand-selected an observer to be within a spiral arm region largely unaffected by any SN feedback for later analysis and comparison. These observer positions are labeled P1, P2, P3, and P4 hereafter and have a distance of $8.16\ \mathrm{kpc}$, $6.50\ \mathrm{kpc}$, $10.22\ \mathrm{kpc}$, and $8.38\ \mathrm{kpc}$, respectively, from the Galactic center (see Fig. \ref{fig:Midplanes}). A detailed description and analysis of the impact of these locations with respect to observations is provided in Sect. \ref{sect:LocalBubble}.

{ In Fig. \ref{fig:Midplanes} we present cuts through to the disk midplane for the selected snapshot of the MHD simulation at $t= 250\ \mathrm{Myr}$. The cut of the gas number density distribution $n_{\mathrm{g}}$ shows the four well-defined spiral arms. The arms blend into a diffuse disk in the center since we are lacking a proper bulge treatment. However, this does not affect the RT modeling and analysis performed in this paper since our focus is on spiral arm detections in the outer regions. The electron distribution $n_{\mathrm{el}}$ has a similar pattern as the gas density with values comparable to other models of the Milky Way \citep[][]{Cordes2002,Yao2017,Pellegrini2019}. However, the electron abundance is lower at the very center of the spirals. The initial magnetic seed field with $0.04\ \mu\mathrm{G}$ has evolved into a configuration following the spiral structure of the gas with a magnitude of about $1-10\ \mu\mathrm{G}$. Such a field strength is what is expected for a Milky Way model \citep[e.g.][]{Beck2001}. The field direction is mostly toroidal but has no strong reversals as claimed to be present in the  magnetic field of the Milky Way \citep[see][and references therein]{Sun2008,Beck2013}. Consequently, some of our followup synthetic synchrotron and $RM$ observations may lack  features such as additional zero transition in the $RM$ signal characteristic for the  Milky Way. The velocity is well behaved with an orbital velocity of $220\ \mathrm{km\ s}^{-1}$ within the disk. This is consistent with the velocity field of the Galactic disk at a distance of roughly $8\ \mathrm{kpc}$ \citep[][]{Bovy2012} where our Solar system and the observers P1-P4, respectively, are located. More recently, \cite{Reid2019} estimated a velocity of $236\pm 7\ \mathrm{km s}^{-1}$. However, a somewhat lower orbital velocity does not substantially effect our synthetic spectral line observations (see Sect. \ref{sect:COLVDiagram}). }

\section{Radiative transfer post-processing}
\label{sect:RTPostProcessing}
In this section we briefly outline the post-processing techniques and the underlying physics of the synthetic observations produced by different tracers. The RT simulations are performed with the RT code \textsc{Polaris}\footnote{\url{http://www1.astrophysik.uni-kiel.de/~polaris/}} \citep[][]{Reissl2016}. The advantage of \textsc{Polaris} is that it unites the physics of dust emission \citep{Reissl2016,Reissl2018a}, line transfer including the Zeeman effect \citep{Brauer2017A,Brauer2017B}, and synchrotron emission \citep{Reissl2019} under a common framework. Furthermore, the code allows one to perform RT simulations on a Voronoi grid. Hence, any \textsc{Arepo} MHD data can be post-processed directly. \textsc{Polaris} RT simulations consider the polarization state of radiation by default. However, within the scope of this paper we focus only on the unpolarized emission of the  different tracers i.e.\ dust, synchrotron, and molecular line emission, respectively, as well as the Faraday rotation measure (RM).

\begin{figure*}
\centering
\begin{minipage}[c]{1.0\linewidth}
     \includegraphics[width=1.0\textwidth]{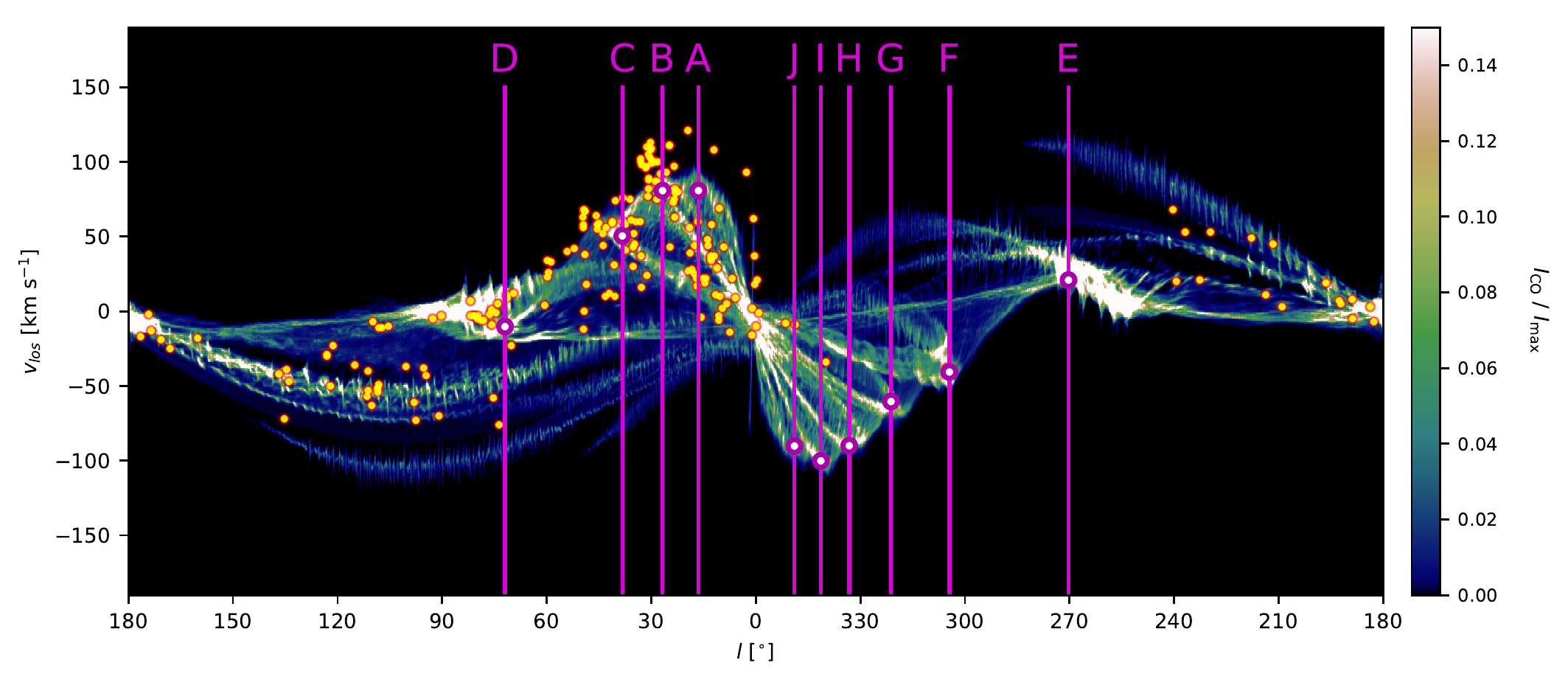}
\end{minipage}

\caption{Synthetic longitude-velocity diagram of the $^{12}\mathrm{CO}$(1-0) emission line constructed using our Galactic model by placing the observer at position P1 (see Fig. \ref{fig:Midplanes}). The velocities are averaged within a latitude of $|b|<\ 1.2^{\circ}$. Purple dots and vertical lines show detections of spiral arm tangent points as determined from CO. Yellow dots represent  
observations of maser emissions coming from high-mass star forming regions within our own Milky Way presented in  \citep{Reid2019}. Note that detections in general depend on the tracers used (see Table \ref{tab:TangentPoints}). The detections are labeled A - J with increasing Galactic longitude. The magnitude of the CO intensity is normalized by its peak value.}
\label{fig:VelMap}
\end{figure*}

\begin{figure*}
\centering
\begin{minipage}[c]{1.0\linewidth}
     \includegraphics[width=0.49\textwidth]{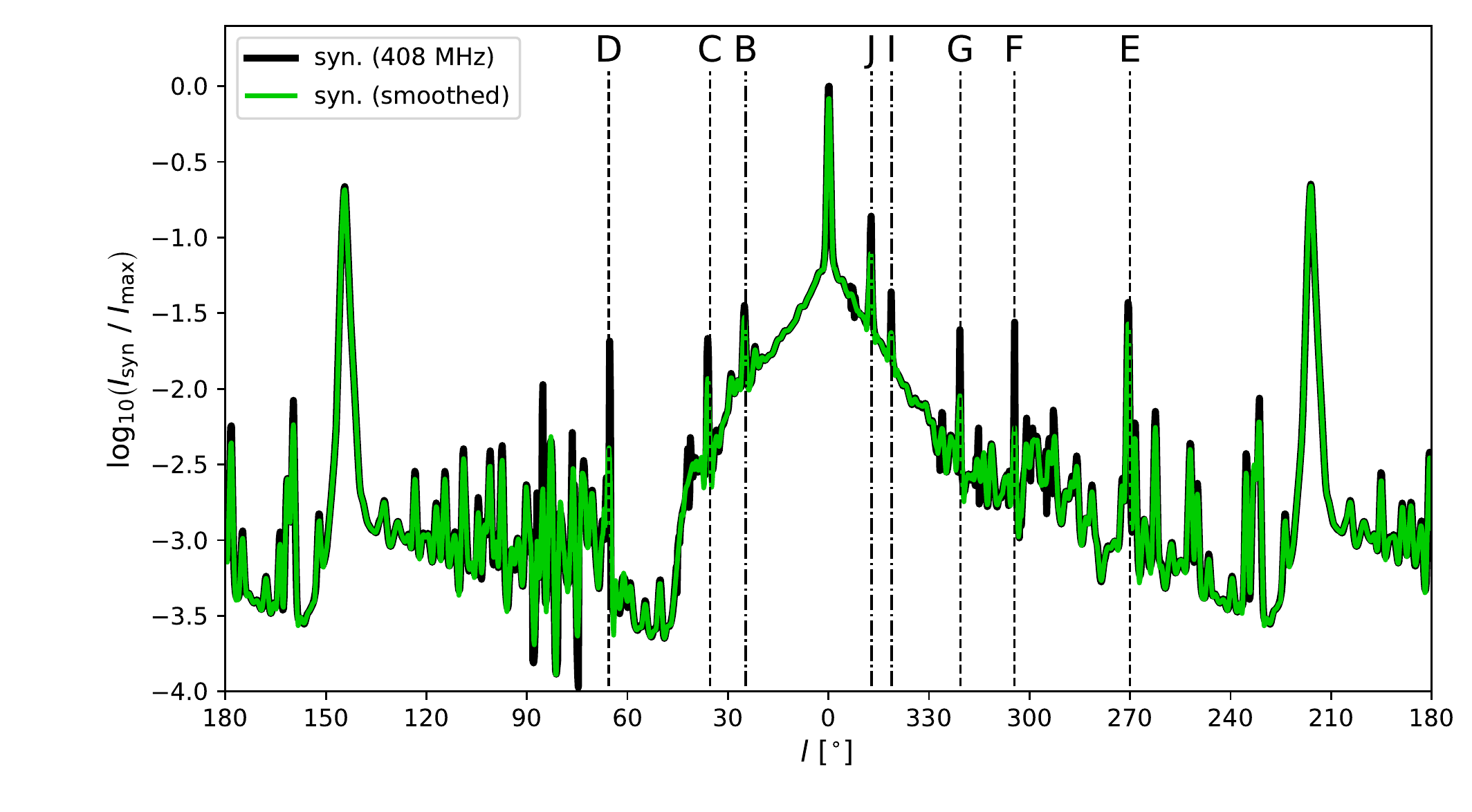}
     \includegraphics[width=0.49\textwidth]{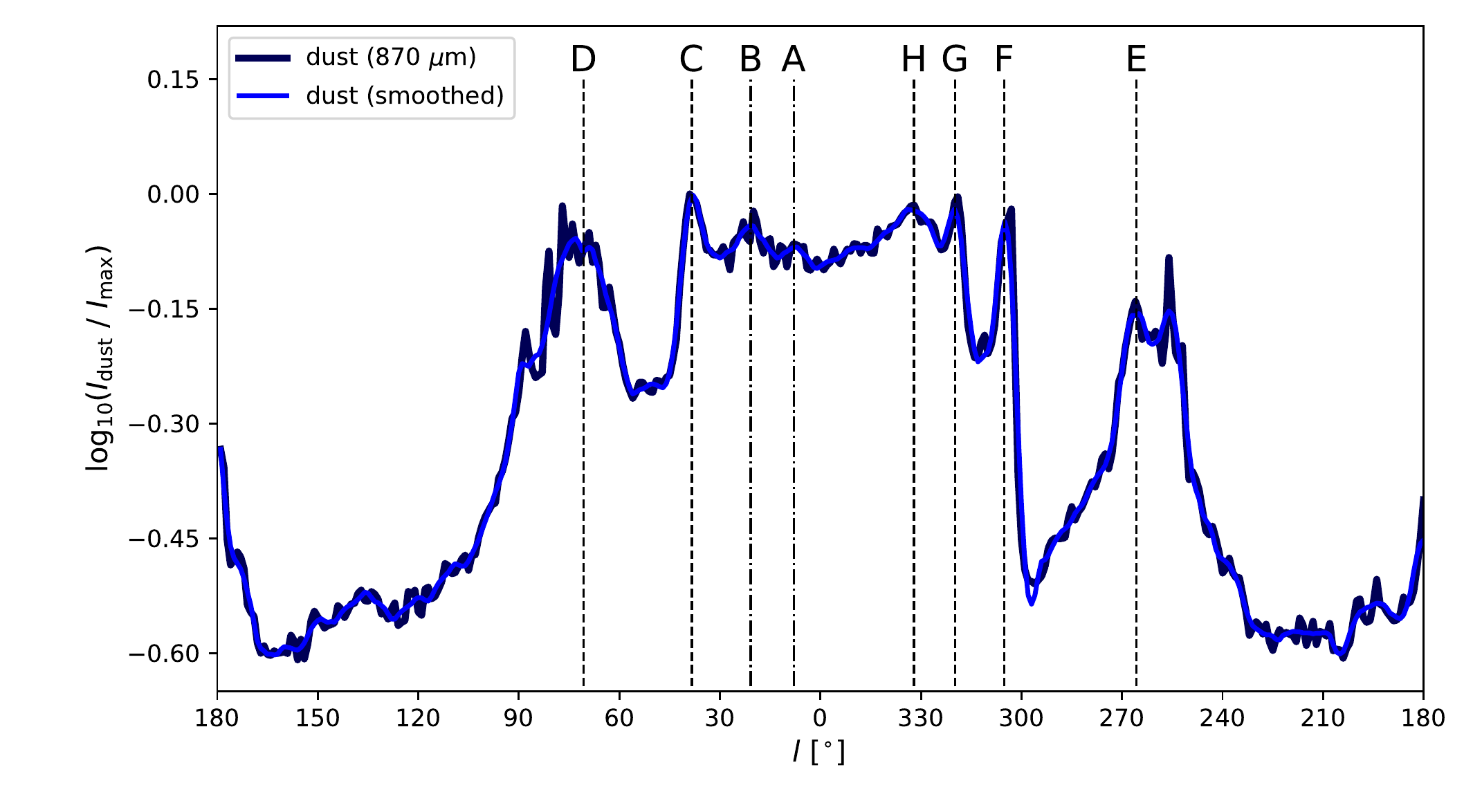}
     
     \caption{Longitudinal profiles of the $408\ \mathrm{MHz}$ synchrotron emission (left) and $870\ \mu\mathrm{m}$ dust emission (right) for our Milky Way model. The panels show  the synchrotron emission averaged profile as well as the dust (blue). The profiles in green and blue are smoothed over $3^\circ$ along the longitude. Vertical black dashed lines and letters indicate tangent points of the spiral arms as detected from each tracer. Note that the longitude of the tangent points are not identical in the left and right panels since the exact position depends on the tracer (see Table \ref{tab:TangentPoints}). All profiles are normalized by their peak values for better comparison and analysis.}
\label{fig:ProfileSyncDust}
\end{minipage}

\begin{minipage}[c]{1.0\linewidth}
      \includegraphics[width=0.49\textwidth]{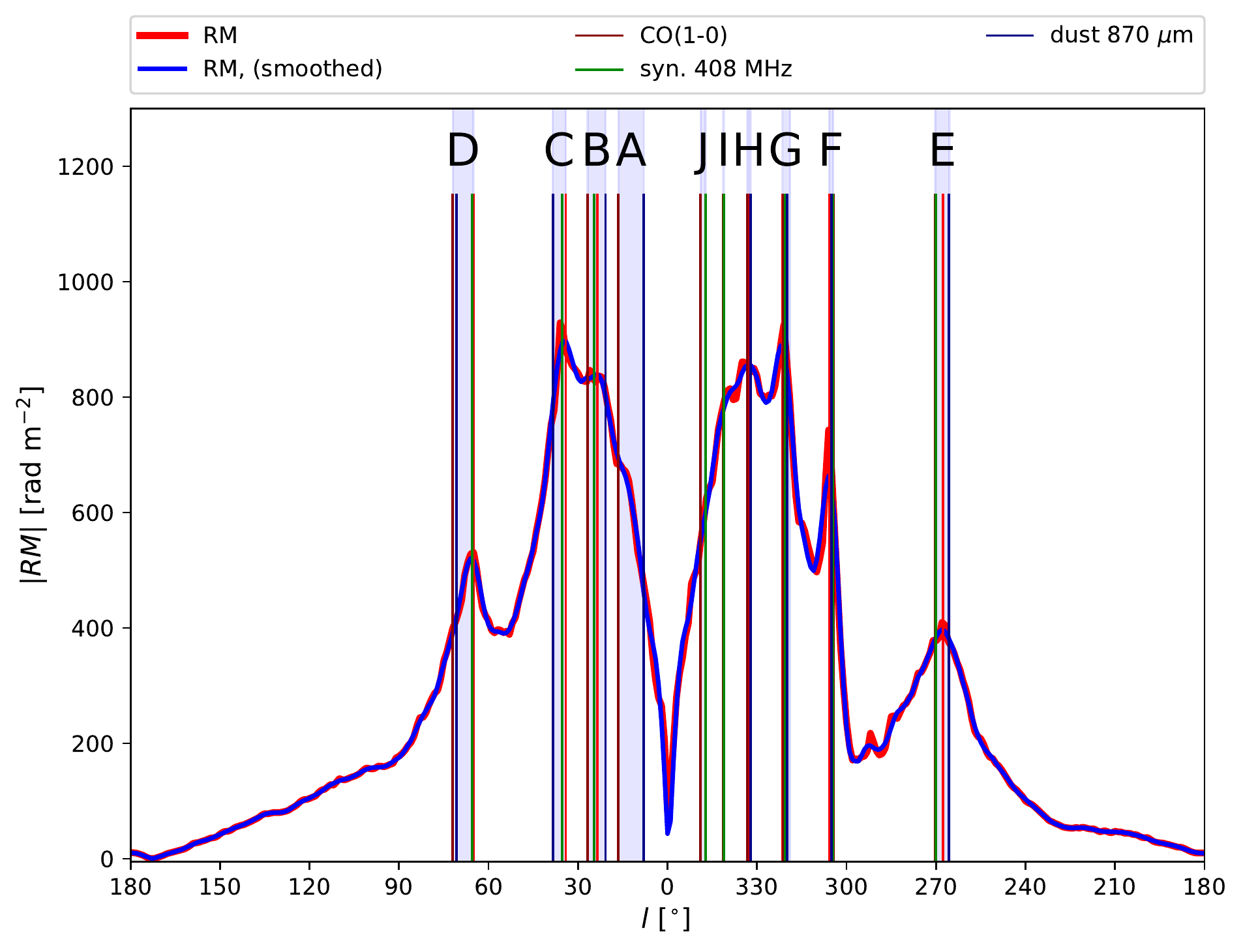}     \includegraphics[width=0.49\textwidth]{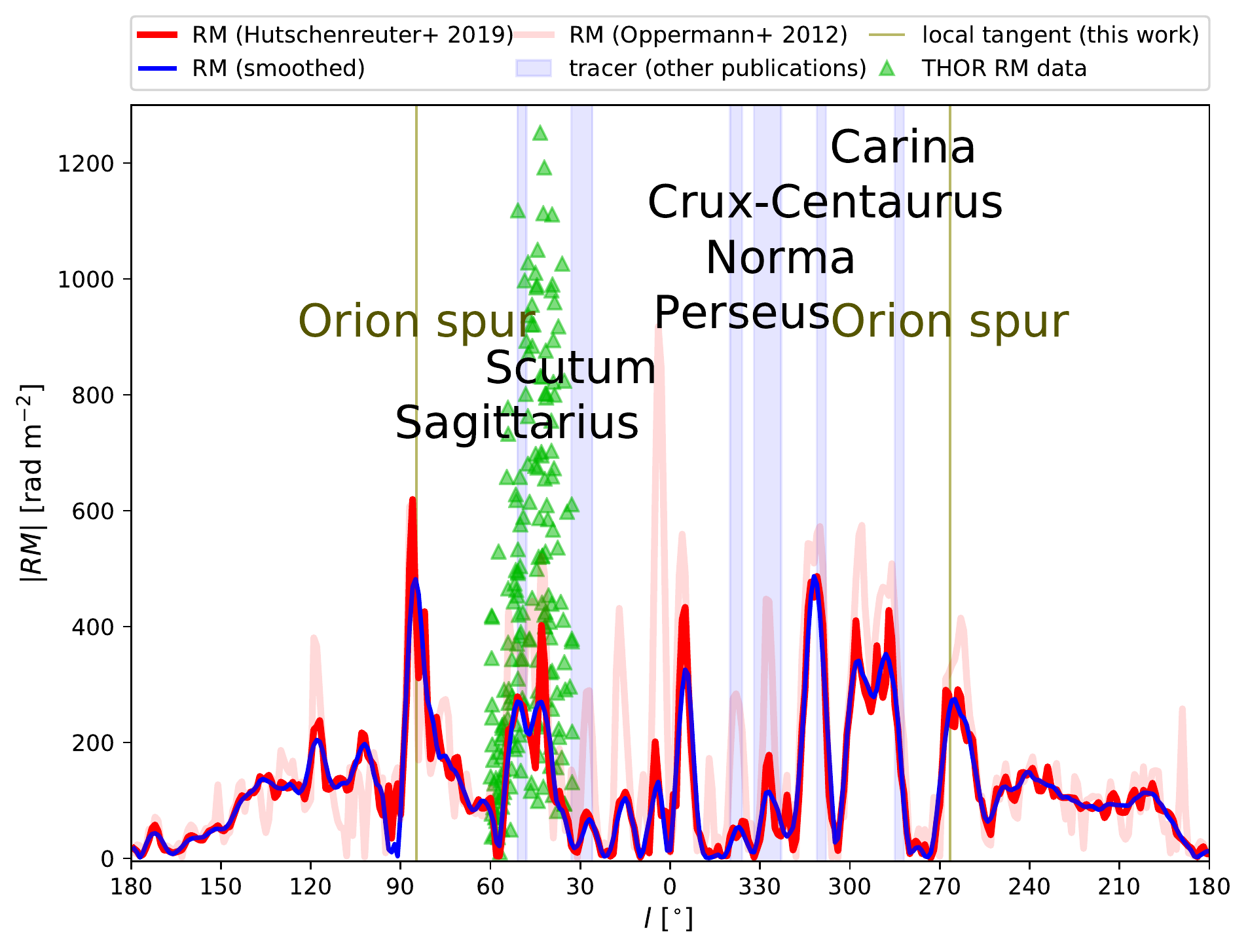}
      
      \caption{{ Left: The synthetic $RM$ profile of our Milky Way model as a function of longitude $l$. Vertical lines represent the tangent point detections as derived from the distinct tracers listed in Table \ref{tab:TangentPoints}. Vertical shaded bars and labels indicate the angular range of the detections for all tracers per spiral arm. Right: The same as the left panel for the observed Milky Way $RM$ \citep{Oppermann2012,Hutschenreuter2019}. Green triangles indicate the THOR $RM$ data presented in \cite{Shanahan2019} as well as newly derived data points. We note that we cut some of the THOR data points beyond $1300\ \mathrm{rad\ m}^{-2}$ for a better comparability with the left panel.}}
\label{fig:ProfileRM}
\end{minipage}
\end{figure*}

\subsection{Dust emission}
For the dust component, we assume the canonical model of the ISM. For this model, the ratio of dust mass to gas mass is taken to be $1\%$. The dust consists of a mixture of materials of $37.5\%$ silicate and $62.5\%$ graphite reproducing the characteristics of the Galactic extinction curve \citep[][]{Weingartner2001}. The grain sizes $a$ range from $a_{\mathrm{min}} = 5\ \mathrm{nm}$ to $a_{\mathrm{max}} = 250\ \mathrm{nm}$ with a size distribution following a power-law of ${ N_{\mathrm{dust}}(a)\propto a^{-3.5} }$ \citep[][]{Dohnanyi1969,Mathis1977}. Since we do not intend to track dust polarization the grains are considered to be spherical for simplicity. We apply absorption cross-sections $C_{\mathrm{abs}}(a)$ per grain size pre-calculated on the basis of laboratory data \citep[see][for a detailed description]{Weingartner2001,Reissl2017}. Assuming equilibrium between absorbed and emitted radiation \citep[][]{Lucy1999,BjorkmanWood2001} we get for the dust emissivity:
\begin{equation}
j_{\mathrm{dust}}=B_{\nu}\left( T_{\mathrm{dust}} \right) \int_{ a_{\mathrm{min}} }^{ a_{\mathrm{max}} } C_{\mathrm{abs}}(a) N_{\mathrm{dust}}(a) \mathrm{d}a\, ,
\end{equation}
where $B_{\nu}\left( T_{\mathrm{dust}} \right)$ is the Planck function and $T_{\rm dust}$ is the dust temperature. The temperature $T_{\rm dust}$  is calculated on-the-fly in the MHD simulation according to the prescription given in \citet{Glover2012}, in which heating by gas-grain energy transfer and the absorption of photons from the ISRF is balanced by dust grain thermal emission. For the purposes of calculating $j_{\rm dust}$ in \textsc{Polaris}, we simply adopt the values of $T_{\rm dust}$ computed in the MHD simulation.

\subsection{Synchrotron radiation and Faraday $RM$}
Synchrotron radiation is emitted from relativistic cosmic-ray (CR) electrons following a power-law distribution in energy ${ N_{\mathrm{el}} \left( \gamma \right) = n_{\mathrm{CR}} \gamma^{-3} }$ where $\gamma$ is the Lorentz factor \citep[][]{Rybicki1979,Bennett2003,Miville2008}. The energy spectrum of the Milky Way is usually taken to be between $\gamma_{\mathrm{min}}=4$ and $\gamma_{\mathrm{max}}=400$ \citep[][]{Webber1998}. 

The CR electrons of the Milky Way follow a smooth distribution. A parametrization of the CR distribution is provided in \cite{Drimmel2001}. However, the scale height and extension of the disk of the MHD simulations is not the same as that of the Milky Way. Hence, we follow the approach discussed in \citet{Reissl2019} and assume equipartition between the CR energy and the magnetic field energy, so that ${ n_{\rm CR} \simeq B^2 / (16\pi \gamma_{\rm min}m_{\rm e}c^2 })$. Here, $B$ is the magnetic field in each cell of the model galaxy, $m_{\rm e}$ is the electron mass, and $c$ is the speed of light. Finally, the synchrotron emissivity is 
\begin{equation}
j_{\mathrm{syn}}= \int_{ \gamma_{\mathrm{min}} }^{ \gamma_{\mathrm{max}} }  P_{\nu}\left( \gamma \right) N_{\mathrm{el}}(\gamma) \mathrm{d}\gamma\, ,
\end{equation}
where $P_{\nu}\left( \gamma \right)$ is the synchrotron power per frequency  $\nu$ \citep[see e.g.][for further details]{Pandya2016,Reissl2019}. We note that the synchrotron power has an angular dependency $P_{\nu}\left( \gamma \right) \propto \sin\vartheta$ where the angle $\vartheta$ is defined to be between magnetic field direction and the line-of-sight (LOS).

In the case of polarized radiation, the polarization angle $\chi_{\mathrm{0}}$ of that radiation rotates while passing through an ionized and magnetized medium \citep[][]{Rybicki1979}. Consequently, the observed orientation angle becomes ${ \chi_{\mathrm{obs}}=\chi_{\mathrm{0}}+\lambda^2 RM }$ \citep[see e.g.][]{Burn1966}. 
The quantity
\begin{equation}
RM = \frac{1}{2\pi}\frac{e^3}{m_{\rm e}^2c^4} \int n_{\rm el} B_{||} d\ell
\label{eq:RM}
\end{equation}
is the wavelength independent rotation measure, $e$ is the electron charge, $n_{\rm el}$ is the density of free electrons, and $B_{||}=B\cos\vartheta$ is the magnitude of the magnetic field component along the LOS \citep[][]{Burn1966,Rybicki1979,Huang2011}. For the $RM$ calculations we utilize the distribution of $n_{\rm el}$ provided by the MHD simulation. We define the $RM$ to be positive for LOSs being parallel with the direction of the magnetic field and vise versa. Thus, the $RM$ provides an estimate of the magnetic field strength as well as the field direction modulated by the distribution of thermal electrons. 

\subsection{Molecular line emission and excitation}
We consider the $^{12}\mathrm{CO}$ density distribution $n_{\mathrm{CO}}$ of the MHD galaxy for the line RT. In rapidly rotating disk galaxies, the difference in velocity between molecular clouds located at different points in the disk is generally larger than the thermal or microturbulent velocities, and so it is generally sufficient to consider only local re-absorption of the emitted photons. This therefore allows us to compute the CO fractional level populations $f_{\rm i}$ using the large velocity gradient method \citep[for details we refer to][]{Ossenkopf1997,Ober2015,Brauer2017B} rather than a full non-LTE treatment. The total emissivity for the transition from level $i$ to $j$ is given by
\begin{equation}
j_{\rm ij}= \frac{h\nu_{\rm ij}}{4\pi} \int_{ \nu_{\mathrm{min}} }^{ \nu_{\mathrm{max}} } n_{\mathrm{CO}} f_{\rm i} A_{\rm ij} \phi(\nu) \mathrm{d}\nu
\end{equation}
where the Einstein coefficient $A_{\rm ij}$ provides the probability of spontaneous emission. For the line broadening we take a Gaussian profile,
\begin{equation}
 \phi(\nu) = \frac{c}{\sqrt{\pi} a_{\mathrm{tot}} \nu_{\rm ij}  } \exp{\left(  -\frac{c^2(\nu-\nu_{\rm ij})^2}{ a_{\mathrm{tot}}^2  \nu_{\rm ij}^2}   \right)}\, ,
\end{equation}
where $\nu_{\mathrm{\rm ij}}$ is the characteristic transition frequency and ${ a_{\mathrm{tot}}^2 = a_{\mathrm{th}}^2 + a_{\mathrm{mtrb}}^2 }$ is the line width parameter where ${a_{\mathrm{th}} = (2 k_{\mathrm{B}} T_{\mathrm{g}} / m_{\mathrm{mol}})^{1/2}  } $ accounts for thermal broadening and $a_{\mathrm{mtrb}}$ for microturbulent velocities.  We estimate the additional broadening of the lines by microturbulence by following the scaling relation ${ a_{\mathrm{mtrb}} \approx 0.52\ \mathrm{km\ s}^{-1} \left( L/\mathrm{pc} \times  \Sigma / 100 \mathrm{M_\odot / pc}^2 \right)^{1/2} }$ suggested in \cite{Heyer2009}. Here, $L=2\left( 3V / 4\pi \right)^{1/3}$ is the characteristic length of a Voronoi cell with volume $V$ and for the surface density $\Sigma$ we take the face-on view of the Milky Way model (see also Fig. \ref{fig:NHMaps}). For the range of parameters provided by our model we get values of ${ a_{\mathrm{mtrb}}=1.4\times 10^{-3} - 4.5\ \mathrm{km\ s}^{-1} }$ within the disk. Indeed, such a range of the microturbulent broadening parameter $a_{\mathrm{mtrb}}$ is comparable to such values that are commonly assumed to be present in the Milky Way \citep[e.g.][]{Smith2002,Shetty2011} justifying the scaling relation approach. We note that the case ${ a_{\mathrm{mtrb}} > 1.0\ \mathrm{km\ s}^{-1} }$ does only occur within the densest regions of the spiral arms and the Galactic center. Hence, we do not expect $a_{\mathrm{mtrb}}$ to have a significant impact on our results, as the total line width of the CO emission from an individual molecular cloud in our simulations is dominated by the effects of resolved cloud-scale motions.

For the \textsc{Polaris} RT simulations we focus on the ${ J = 1 \longrightarrow 0 }$ transition of $^{12}\mathrm{CO}$
since this molecule is commonly used to trace the ISM mass content as well as the Galactic velocity field \citep[e.g.][]{Dame1987,Dame2001,Jackson2006,Sormani2018}. The characteristic parameters, i.e.\ the mass of a CO molecule is $m_{\mathrm{mol}} = 4.65\times 10^{-23}\ \mathrm{g}$, the Einstein coefficient $A_{\mathrm{ij}}=7.20\times 10^{-8}\ \mathrm{s^{-1}}$, and the transition frequency $\nu_{\mathrm{ij}} = 115.27\ \mathrm{GHz}$ are taken from the LAMDA\footnote{\url{https://home.strw.leidenuniv.nl/~moldata/CO.html}} molecular database \citep[][]{Schoier2005}, which is also our source for the collisonal excitation and de-excitation rates \citep{Flower2001,Schoier2005,Jankowski2005,Wernli2006}. We assume that it is only necessary to account for CO-H$_{2}$ collisions, since almost all of the CO in our MHD simulation is located in gas with a high H$_{2}$ fraction.

\subsection{The ray-tracing scheme}
We simulate all-sky maps for the individual tracers applying the \textsc{Healpix}\footnote{\url{https://healpix.jpl.nasa.gov/}} pixelation scheme \citep[][]{Gorski2005}. The number of \textsc{Healpix} sides is taken to be $N_{\mathrm{side}}=256$. This results in a total number of $786432$ pixels on a sphere larger than the MHD grid, corresponding to an angular resolution of $13.7\ \mathrm{arcmin}$. Each pixel on the \textsc{Healpix}-sphere defines a LOS between the pixel itself and the distinct pre-defined observer positions labeled P1, P2, P3, and P4, respectively. For each of the LOSs we perform a ray-tracing by solving the RT problem as outlined above for each individual tracer. Here, \textsc{Polaris} utilizes a Runge-Kutta-Fehlberg solver with built-in error correction working within an error limit of $<\ 10^{-6}$ for each Voronoi cell \citep[see e.g.][for details]{Ober2015,Reissl2019}.

\section{Detection of the Galactic spiral structure}
\label{sect:SpiralDetection}
In this section we present our synthetic observations and evaluate the tangent points as detected using different tracers. Such tangent points manifest themselves along the Galactic longitude axis as characteristic peaks and are usually defined to represent the point where the LOS has exactly one intersection with the spine of a spiral arm. We discuss our results in the context of actual tangent points observed in the Milky Way. For the time being we focus on the observer position P1 representing the most similar conditions to our own environment within the Milky Way. Statistics for the positions P2 - P4 are provided later in Sect. \ref{sect:LocalBubble}.

\subsection{CO longitude-velocity diagram}
\label{sect:COLVDiagram}
{ In Fig. \ref{fig:VelMap} we show the longitude-velocity diagram derived from the synthetic CO all-sky observations as seen from the observer position P1 within the latitudinal range of $|b| <  1.2^{\circ}$. In detail, we select all pixels within that particular range of $|b|$ of the \textsc{Healpix} sphere along the entire longitude $l$ per velocity channel $v_{\mathrm{los}}$ and count the values within a certain bin with $(l,v_{\mathrm{los}})$ as coordinates. Furthermore a latitudinal range of $|b| <  1.2^{\circ}$ is chosen to be comparable to the range of latitudes observed in the THOR survey \citep[see][]{Beuther2016,Wang2020}.} { We assume the observer to be co-moving with the Galactic disk rotation with a local velocity of $v_{\mathrm{g}}=220\ \mathrm{km\ s}^{-1}$ at P1}. The resulting features are very similar to CO observations of the  Milky Way \citep{Dame1987, Dame2001} as well as the HI emission map shown in \cite{Reid2019}. For a better comparison between our Galactic model and the Milky Way we also plotted the maser measurements in Fig. \ref{fig:VelMap} as listed in \cite{Reid2019}. This comparison reveals that our model underestimates the size of the velocity features towards the galactic center. However, our Milky Way model does not include the Galactic bar which is responsible for the high-velocity peaks in the longitude-velocity plots of the central regions \citep[e.g.][]{Sormani2015c}. Furthermore, our overall orbital velocity is somewhat lower than the one suggested in \cite{Reid2019}.

The diagram exhibits ten well-defined loops that are the projection of the spiral arms in the longitude-velocity plane. Tangent points are identified by the reversals of the loops \citep[see e.g][]{Dame2001,Sormani2015c,Reid2019}. The tangent point candidates detected according to this criterion are labeled in alphabetical order with increasing latitude by letters A - J for further discussions. 

\begin{figure}
\centering
\begin{minipage}[c]{1.0\linewidth}
     \includegraphics[width=1.0\textwidth]{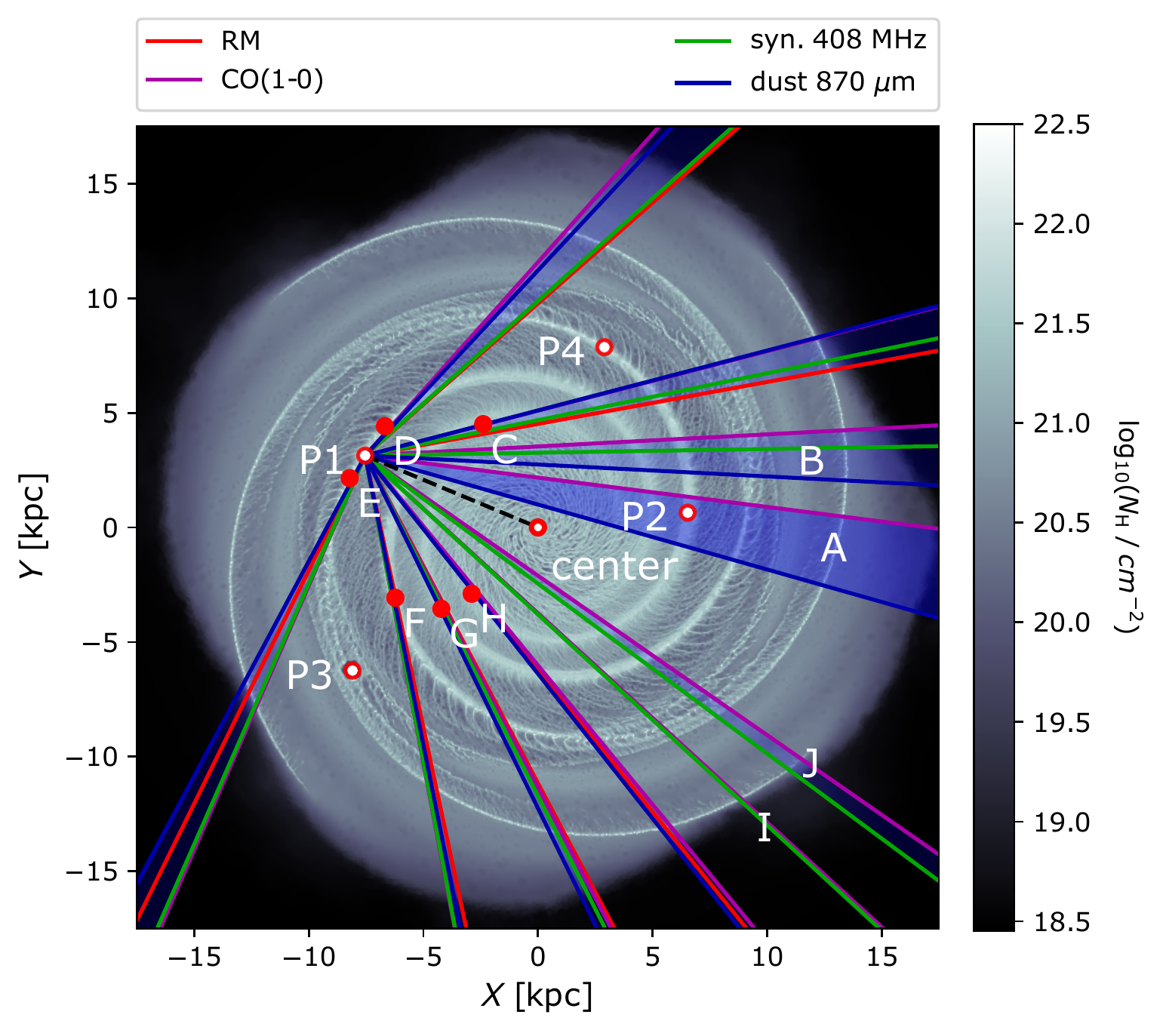}
\end{minipage}

\caption{Face-on column density map of our Milky Way model. Red circles are the observer positions and the Galactic center, respectively, while red dots represent the tangent points. The black dashed line represents the LOS towards the center i.e. l $=0^{\circ}$ in the Galactic coordinate system. Colored lines indicate the detected spiral arms for different tracers and the blue shaded areas show the angular range of detections for each spiral arm.}
\label{fig:NHMaps}
\end{figure}

\subsection{Radio observations}
A model of the Milky Way was presented in \cite{Beuermann1985} based on the $408\ \mathrm{MHz}$ all-sky radio observation of \cite{Haslam1981A,Haslam1982B}. This model provides an estimate for the Milky Way's spiral arms. We create a similar all-sky map of the $408\ \mathrm{MHz}$ synchrotron radiation with the help of \textsc{Polaris}, which delivers synchrotron emission and $RM$ simultaneously in a single RT simulation \citep[see][for details]{Reissl2019}. { Then we average the synchrotron signal within latitude $|b|<\ 0.2^{\circ}$ and plot it as a function of Galactic longitude. Here, we decrease the latitudinal range compared to CO in order for the spiral arm to become distinguishable from the synchrotron signal. For a larger range of $|b|$ the signal becomes too noisy for any detection. Furthermore, we smooth the resulting profile along $l$ to get a well defined maximum for each tangent point detection. 

We report that the peaks in our model are much narrower than the peaks presented in \cite{Beuermann1985}. This is most likely due to our equipartition assumption implying that the CR electron distribution $n_{\mathrm{CR}}$ only depends on $B^2$ and so CR electrons are mostly present close to the spines of the spiral arms. Hence, we have a scale height that is somewhat smaller than that of the Milky Way and a synchrotron signal may not be detectable from noise for a larger range of $|b|$. A more sophisticated CR propagation scheme may also provide broader peaks comparable to those of \cite{Beuermann1985}. Nevertheless, the location of the tangent points can be estimated from the well-defined peaks in the synchrotron signal that are within the orbital cycle of P1 and above the local noise level. The resulting profiles are shown in Fig. \ref{fig:ProfileSyncDust}. }

Here, we find that the tangent points A and H are not detectable in the synchrotron signal. We speculate that this may be due to the $\sin\vartheta$ dependency of the synchrotron signal. The LOS and the magnetic field directions are not strictly parallel for most of the tangent points, i.e.\ the angular dependency of the synchrotron emission is $\sin\vartheta>0$. Hence, the strong increase in the magnetic field strength can compensate for the decreasing factor coming from the angular dependency and most tangent arms can be detected by synchrotron observations. However, this is not the case for the tangent point H where $\sin\vartheta\approx 0$. 

\subsection{Cold dust signatures}
Tangent points may also be detected in emission from cold dust. Corresponding dust observations of $240\ \mu\mathrm{m}$ and $870\ \mu\mathrm{m}$ emission are provided in \cite{Drimmel2000} and \cite{Beuther2012}, respectively. However, \cite{Beuther2012} counted the dust emission of distinct sources, whereas we take the profile of the continuum emission along the entire galactic plane. The procedure of detecting the tangent points along the longitude is similar to that of the synchrotron detections. Again, we consider the range of $|b|\ < 1.2^{\circ}$ as in the THOR survey \citep[][]{Beuther2016,Wang2020}.

In Fig. \ref{fig:ProfileSyncDust} we present the profile of the $870\ \mu\mathrm{m}$ emission. In contrast to \cite{Beuther2012}, we do not find a central peak. This may be because our MHD simulations lack a proper bar/bulge model. However, the profile has some well-defined local maxima that can be clearly associated with the tangent points of the spiral arm candidates A - H. 
The synthetic profile of the $240\ \mu\mathrm{m}$ dust emission is virtually identical to the $870\ \mu\mathrm{m}$ profile for our Milky Way model. Hence, we discuss only the $870\ \mu\mathrm{m}$ dust observations in the following sections.
\begin{table*}
\centering
   \begin{tabular}{|C{1.4cm}| S	| S | S | S | S | S | S | S | S | S | S |}
    \toprule
      \multicolumn{1}{|c|}{	Tangent point				} &
      \multicolumn{1}{c|}{	CO(1-0)             } &
      \multicolumn{2}{c|}{	$\rm N_H$			} &
      \multicolumn{2}{c|}{	RM					} &
      \multicolumn{2}{c|}{	Syn. 408 MHz			} &
      \multicolumn{2}{c|}{ 	Dust 870 $\mu{\rm m}$ 	} \\
            						& 
      l$\, [^\circ]$ 					& 
      l$\, [^\circ]$ & $\Delta $l$\, [^\circ]$ 	& 
      l$\, [^\circ]$ & $\Delta $l$\, [^\circ]$ 	& 
      l$\, [^\circ]$ & $\Delta $l$\, [^\circ]$ 	& 
      l$\, [^\circ]$ & $\Delta $l$\, [^\circ]$  \\
      \midrule
A  &  16.46   &  X       &  X      &  X       &  X      &  X        &  X      &  7.91    &  8.55   \\
B  &  26.76   &  X       &  X      &  23.47   &  3.29   &  24.68    &  +2.08  &  20.76   &  +6.00  \\
C  &  38.30   &  38.38   &  -0.08  &  34.11   &  +4.19  &  35.31    &  +2.99  &  38.38   &  -0.08  \\
D  &  71.97   &  74.45   &  -2.48  &  65.08   &  +6.89  &  65.56    &  +6.41  &  70.72   &  +1.25  \\
E  &  270.30  &  264.01  &  +6.29  &  267.88  &  +2.42  &  270.00   &  +0.30  &  265.69  &  +4.61  \\
F  &  304.46  &  304.30  &  +0.16  &  305.81  &  -1.35  &  304.52   &  -0.06  &  305.11  &  -0.65  \\
G  &  321.21  &  319.48  &  +1.73  &  321.53  &  -0.32  &  320.65   &  +0.56  &  319.83  &  +1.38  \\
H  &  333.20  &  329.75  &  +3.45  &  332.62  &  +0.58   &  X        &  X      &  332.11  &  +1.09  \\
I  &  341.35  &  X       &  X      &  X       &  X      &  341.13   &  0.22   &  X       &  X      \\
J  &  348.91  &  X       &  X      &  X       &  X      &  347.18   &  1.73   &  X       &  X      \\
\bottomrule
  \end{tabular}
  \caption{{ Tangent points of spiral arms in our Milky Way model as determined using different tracers as seen from the example observer position P1. The longitude of the tangent point is indicated by l, while $\Delta$l gives the difference with respect to the values determined from $\rm CO(1\to0)$ (see Fig. \ref{fig:VelMap}). The X marks tangents with no detection. We note that the tabulated angles merely represent our Milky Way analog and are not to be directly compared with actual observations of the Milky Way.}}
  \label{tab:TangentPoints}
\end{table*}
\subsection{Longitudinal Faraday $RM$ profiles}
Faraday rotation is sensitive to the magnetic field strength $B$, electron fraction, and the direction of the field lines with respect to the LOS (i.e., $\cos\vartheta$). Within the spiral arms the field strength $B$ reaches a maximum and $\cos\vartheta\approx \pm 1$ (see Fig. \ref{fig:Midplanes}). Consequently, we expect the $RM$ profile to have well-defined features corresponding to the tangent points coming mostly from the magnetic field properties. However, this may only be true for our model and might not match the situation in the Milky Way arms themselves, where an enhanced electron density may also contribute to the $RM$ signal \citep{Shanahan2019}.

\cite{Oppermann2012} produced an all-sky map of Galactic Faraday rotation using all available $RM$ data. This all-sky map was later improved by taking the free-free emission into account \citep[][]{Hutschenreuter2019}. 

Conveniently, the $RM$ of  \mbox{\cite{Oppermann2012}}  and  \citep[][]{Hutschenreuter2019} are publicly available in the \textsc{Healpix} 
format\footnote{Oppermann map:\\ \url{wwwmpa.mpa-garching.mpg.de/ift/faraday/}\\Improved Hutschenreuter map:\\ \url{wwwmpa.mpa-garching.mpg.de/~ensslin/research/data/faraday_revisited.html}}. As for the other tracers we create longitudinal profiles by averaging all pixels within the range of $|b|\ < 1.2^{\circ}$. 

In Fig. \ref{fig:ProfileRM} we compare synthetic $RM$ profiles (left panel) with Galactic $RM$s derived by \cite{Oppermann2012} and \cite{Hutschenreuter2019} (right panel). The $RM$ profiles in Fig. \ref{fig:ProfileRM} are shown in comparison with the set of tangents derived from the other tracers and the observed Milky Way arm tangent points as presented in \cite{Beuermann1985}, \cite{Taylor1993}, \cite{Drimmel2000}, \cite{Beuther2016}, \cite{HouHan2014}, and \cite{HouHan2015}, respectively, and the references therein. 

In contrast to the observed $RM$ profile from \cite{Hutschenreuter2019} (right panel), our synthetic map (left panel) displays an increase in the magnitude of the Faraday $RM$ towards the center. This is a result of the strong increase in the magnetic field strength towards the center of our Milky Way model galaxy since we are not properly modeling the bulge region. We also note that our magnetic field is mostly toroidal lacking reversals of the field lines within the plane of the disk as predicted by other Galactic models \citep[e.g.][]{Jansson2012,Grand2017}. Hence, we are missing some of the structure coming from such reversals in our $RM$ profile. The minimum of the profile at l $=0^{\circ}$ indicates the change in sign of the $RM$ where the magnetic field direction goes from parallel ($\cos\vartheta>0$) to anti-parallel ($\cos\vartheta< 0$) with respect to the LOS (see Eq. \ref{eq:RM} and Fig. \ref{fig:Midplanes}). 
The synthetic $RM$ profile in Fig. \ref{fig:ProfileRM} has well defined maxima. Assuming that these maxima correspond to the spiral arm tangents, we detect the candidates B-H.

{ The Galactic $RM$s of \cite{Oppermann2012} and \cite{Hutschenreuter2019} also show peaks in the Galactic plane along the directions of known spiral arms. The width of these peaks in longitude varies significantly, and it does not appear to be correlated with the longitude extent of spiral arm tracers for each arm. The height of the peaks varies from arm to arm, while some $RM$ peaks of similar magnitude appear not related to any known arm. A detailed discussion of each spiral arm tangent is beyond the scope of this paper. However, we make two comments that are significant when comparing the left and right panels of Fig.~\ref{fig:ProfileRM}. }

{ First, the magnitude of the fluctuations in the $RM$s of \cite{Oppermann2012} and \cite{Hutschenreuter2019} is of the same order of magnitude as $RM$s from individual HII regions \citep[see e.g.][for a discussion]{Shanahan2019}. It is therefore possible that some observed $RM$ peaks are related to individual HII regions along the LOS. These are not modeled in the present simulation. }

{ Second, \cite{Shanahan2019} showed that some of the surveys included by \cite{Oppermann2012} and \cite{Hutschenreuter2019}, respectively, were biased against very high $|RM|$. This bias was introduced by the use of data from the NRAO VLA Sky Survey (NVSS), which is limited to $|RM|\ \lesssim 400\ \mathrm{rad\ m}^{-2}$, \citep[in the case of][]{Taylor2009}, or by selection of polarized sources from the NVSS source catalog for new observations. The recent survey by \citet{Schnitzeler2019} explicitly searched for targets with $|RM|\ < 1000\ \mathrm{rad\ m}^{-2}$. The Bayesian analysis cannot correct a bias in the data, if it exists, as in the case of the Sagittarius arm.}

\section{Analysis of arm properties}
\label{sect:ArmProperties}
In this section we discuss the characteristic profile shapes associated with tangent points and a method to recover the pitch angles of the spiral arms on the basis of the tangent points. 

\subsection{The Milky Way from an outside perspective}
\label{sect:OutsideObserver}
In addition to the all-sky maps, we create a face-on map of the column density $N_{\mathrm{H}}$ as seen from an outside observer. In Fig. \ref{fig:NHMaps} we show the column density map and depict the tangent points detected from the perspective of observer P1. For reference, in this section we additionally consider the tangent points as determined from the maxima in the longitudinal profile of $N_{\mathrm{H}}$. All tangent points of the different tracers are listed in Table  \ref{tab:TangentPoints} for the observer P1 (the statistic of the other observer positions is discussed in Sect. \ref{sect:LocalBubble}). Comparing the tangent points as detected by different tracers, we find a maximal scatter of about $\pm 6^{\circ}$, which is comparable to the range e.g. presented in \cite{HouHan2014,HouHan2015} for the Milky Way. Comparing Figure \ref{fig:NHMaps} with Figures \ref{fig:ProfileSyncDust} and \ref{fig:ProfileRM}, it is apparent that the spiral arm tangents closer to P1 are broader in longitude, e.g. D and E, while the most distant spiral arm tangents, in the inner Galaxy, appear to blend together, e.g. A and B, or H, I, and G. The perspective of the observer is important in the sense that spiral arm tangents near longitudes $\pm 45^\circ$ display the highest contrast.

Comparing the tangent points as detected from the embedded perspective of observer P1 (defined as peaks in longitudinal profiles of various tracers) with the face-on map reveals that only the points C, D, E, F, G and H are actually associated with clearly defined spiral arms. The tangents A, B, I, and J are associated with blurred arms that are blending into the disk of diffuse material near the Galactic center in the MHD simulation. Here, we remind the reader that in our Milky Way model we are missing a proper bar and bulge as e.g. provided in the numerical simulations of \cite{Sormani2015a}. However, such simulations do not include a magnetic field component. 

{ The arm structure of the Milky Way outside the central bar region can approximately be described by logarithmic spirals \citep[][]{Vallee2015,HouHan2015}. This is also the model we adopt in this study whereas the log-periodic spirals suggested in \cite{Reid2014} would also account for a possible kink within a spiral arm. Following  trigonometrical considerations the pitch angle $\Psi$ of any arm may be estimated by 
\begin{equation}
\Psi=\tan^{-1} \left[   \frac{\ln\left( \sin(l_{\mathrm{180}})/\sin(2\pi-l_{\mathrm{360}})\right)}{(l_{\mathrm{180}}-l_{\mathrm{360}}+\pi)} \right]\, .
\label{eq:Pitch}
\end{equation}
Here, the quantity $l_{\mathrm{180}}$ is the longitude of a tangent point between $0^\circ$ and $180^\circ$ where as $l_{\mathrm{360}}$ is a tangent between  $180^\circ$ and $360^\circ$. Consequently, the two tangent points $l_{\mathrm{180}}$ and $l_{\mathrm{360}}$ are required to belong to the same spiral arm, which runs from one tangent point to the other.} For example, the tangent points C and F belong to the same arm called $ARM_{\mathrm{CF}}$, hereafter, and D and E constitute the tangent points of the arm $ARM_{\mathrm{DE}}$. Note that $ARM_{\mathrm{DE}}$ is the local arm of the observer position P1. The arms corresponding to the tangent points G and H possess no second tangent point, similarly to the Perseus and Norma arms in our own Galaxy (see Fig. \ref{fig:Sketch}). 

\begin{table}
\centering
   \begin{tabular}{|C{2.3cm}|C{2.3cm}|C{2.3cm}|}
    \toprule
  Tracer   & $ARM_{\mathrm{CF}}$ & $ARM_{\mathrm{DE}}$\\
      \midrule

    $N_{\mathrm{H}}$& $10.78^{\circ}$ & $10.80^{\circ}$ \\
    CO (1-0)& $10.75^{\circ}$ & $8.94^{\circ}$ \\
    $RM$ & $12.98^{\circ}$ & $13.70^{\circ}$ \\
    Syn. $408\ \mathrm{MHz}$& $12.83^{\circ}$ & $12.41^{\circ}$ \\
    Dust $870\ \mu\mathrm{m}$& $10.33^{\circ}$ & $11.86^{\circ}$ \\
       \midrule
     mean $\pm$ STD & $11.72^{\circ} \pm 1.38^{\circ}$ & $11.73^{\circ} \pm 2.01^{\circ}$\\

\bottomrule
  \end{tabular}
  \caption{Pitch angle of the spiral arms $ARM_{\mathrm{CF}}$ and $ARM_{\mathrm{DE}}$ for the different considered tracers as seen from the observer position P1. Here, the arm $ARM_{\mathrm{CF}}$ is defined by the tangent points C and F where as $ARM_{\mathrm{DE}}$ is the local arm of observer P1 with tangent points D and E. Mean values and the standard deviations (STD) are calculated without the angles of $N_{\mathrm{H}}$.}
  \label{tab:PitchAngles}
\end{table}

\begin{figure*}
\centering
\begin{minipage}[c]{1.0\linewidth}
     \includegraphics[width=1.0\textwidth]{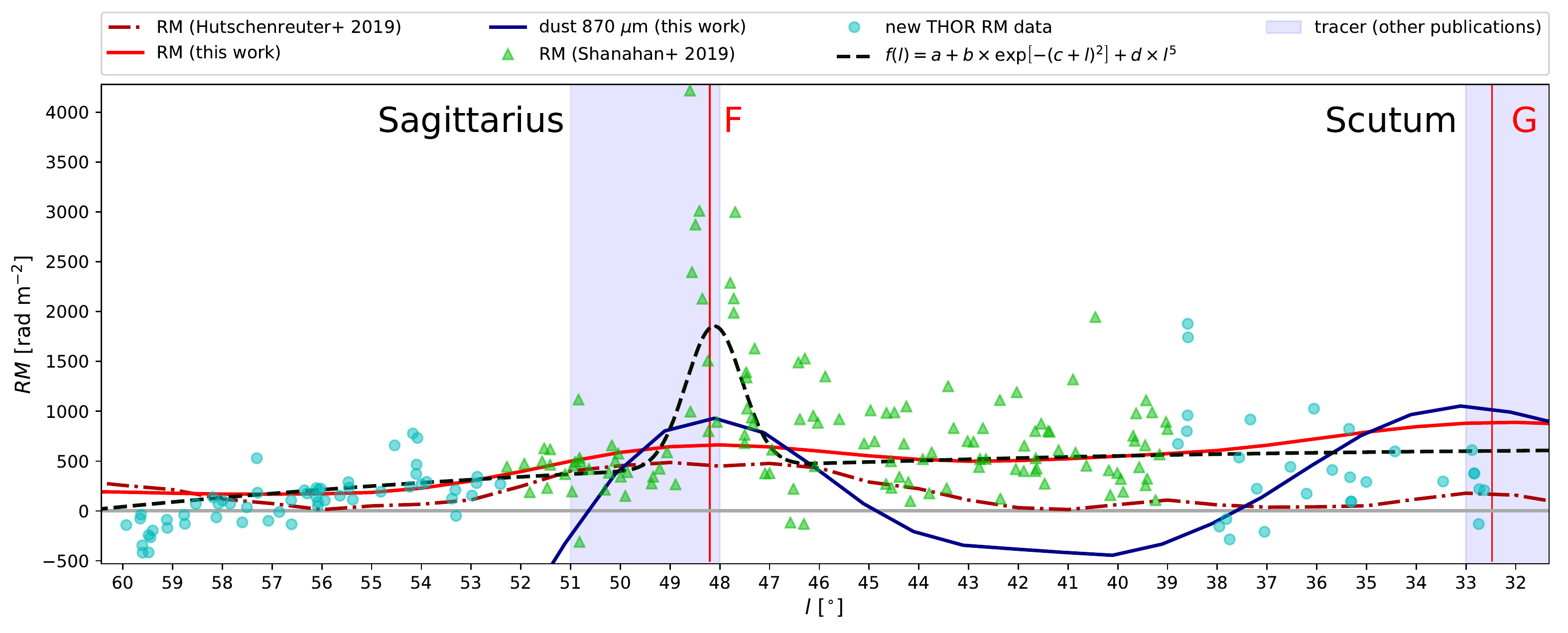}
\end{minipage}

\caption{ Synthetic and measured $RM$ values. The synthetic $RM$ (solid red) and dust (solid blue) are compared to the $RM$ observations  (dashed dotted brown) derived from the \cite{Hutschenreuter2019} map, and the THOR survey, respectively. The THOR data contains measurements (green triangles) already published in \cite{Shanahan2019} as well as newly calculated $RM$ (green dots). All synthetic profiles are shifted and mirrored to match with their tangent point F, which is the peak value of the \cite{Shanahan2019} data at $l = 48.6^\circ$ in the Galactic coordinate system. For comparison, the synthetic dust profile shown in Fig. \ref{fig:ProfileSyncDust} is magnified to match the $RM$ data. The parameters of the function $f(l)$ are fitted to the full THOR data set (dashed black). Vertical red lines indicate the tangent points F and G of the synthetic $RM$ profile shown in Fig. \ref{fig:ProfileRM} whereas vertical blue bars indicate the range of tangent points of different tracers for the Sagittarius and the Scutum arm of the Milky Way.}
\label{fig:ProfileRMComp}
\end{figure*}

The pitch angles $\Psi$ corresponding to the various spiral arms in our Milky Way model are listed in Table \ref{tab:PitchAngles}. All these values are determined using Eq. \ref{eq:Pitch}, with the exception of $N_{\mathrm{H}}$ which is determined from the face-on map. The arm $ARM_{\mathrm{CF}}$ has an average pitch angle of $\Psi = 11.72^{\circ}$ and the local arm $ARM_{\mathrm{DE}}$ has $\Psi = 11.73^{\circ}$. These are average values calculated from various tracers but without the contribution of the $N_{\mathrm{H}}$ data, which has to be excluded since these angles are not derived from the perspective of an inside observer. The recovered pitch angles of $ARM_{\mathrm{CF}}$ and $ARM_{\mathrm{DE}}$ fall still well within to the range of pitch angles of ${ 7^{\circ} \lessapprox \Psi \lessapprox 15^{\circ} }$ reported for the spiral arms of the Milky Way \citep[e.g.][]{Taylor2009,HouHan2014,Yao2017,Chen2019,Reid2019}. Similar values of $\Psi$ are found for other spiral galaxies e.g. $\Psi< 25^{\circ}$ for M83 \citep[][]{Frick2016}.

The $\Psi$ values determined from the perspective of an embedded observer are very close to the values determined from the face-on $N_{\rm H}$ map. This validates the method provided by Eq. \ref{eq:Pitch} to estimate the pitch angle of spiral arms.  However, we note that the systematic offset illustrated by our geometrical model presented in Sect. \ref{sect:SpiralArmProfile} has a significant effect on the recovered value of the pitch angle. For any inner arm border close to the center we find a pitch of $\Psi \approx 12^\circ$, whereas for the outer arm border, we find values closer to $\Psi \approx 9^\circ$. This is not a dramatic error considering that the tracers do indeed probe different regions within a spiral arm. The derived angles $\Psi$ presented in Table \ref{tab:PitchAngles} fall well within the actual observed range of $9^\circ < \Psi < 12^\circ$. However, it illustrates the importance of taking into account the systematic geometric effect explained in Fig. \ref{fig:Sketch} and the difference between the various tracers when determining the pitch angle of the Milky Way spiral arms using Eq. \ref{eq:Pitch}.

Furthermore, for $ARM_{\mathrm{CF}}$ and $ARM_{\mathrm{DE}}$, the pitch angle of the gaseous spiral arms are close to the pitch angle of $\Psi = 15^\circ$ applied in the gravitational potential of the MHD simulation (see Sect. \ref{sect:SimSetup}). However, they are not identical -- the pitch angle of the gas is smaller than the pitch angle $\Psi = 15^{\circ}$  of the applied spiral potential (see Sect. \ref{sect:NumMethods}). It was noted in \citet{Roberts1969} and \citet{Sormani2017} that the gas may shock in front of or even behind the minimum of the potential well. A similar explanation for an observed offset between gas content and star-formation tracers in the Milky Way is discussed in \cite{Ragan2018} and the references therein in greater detail. The exact position of the shock heavily depends on the local parameters such as the sound speed.
{ Indeed, in Fig. \ref{fig:Midplanes} we see that the plotted spirals indicating the potential are not always congruent with the density maximum of the spiral arms. We speculate that this offset between the potential minimum and the shock may account for the deviation between the pitch angle present in the applied potential and the pitch angles recovered from the observational tracers.}

\subsection{The pitch angle of the Milky Way's local arm}
\label{sect:OrionSpur}
{ In the previous section, we showed that the pitch angle of the local arm may be derived from the synthetic $RM$ profile. Indeed, we note two particular features in the $RM$ profile of the Milky Way at $84.68^{\circ}$ and $266.61^{\circ}$, respectively. We marked the features in the right panel in Fig. \ref{fig:ProfileRM} for a better comparison). These features, are clearly within the Solar orbit and we speculate that they originate from our own local arm, the Orion spur (see Fig. \ref{fig:Sketch}).} Consequently, both features would belong to the same spiral arm and we can apply Eq. \eqref{eq:Pitch} to determine its pitch angle. This would result in a pitch angle of $\Psi=4.36^{\circ}$ for the Orion spur. 

This finding is well within the range of pitch angles proposed in the literature by studies utilizing various tracer techniques. In \cite{Yao2017} a pitch angle of $\Psi=2.77^{\circ}$ is derived by modeling the free electrons in the Milky Way, \cite{Chen2019} report a value of $\Psi=10.20^{\circ}$ based on the distribution of observed O/B-type stars, while \cite{Reid2019} estimate an angle of $\Psi=11.40^{\circ}$ by observing molecular masers of young high-mass stars. 
Hence, the features in the $RM$ signal of the \cite{Oppermann2012} and \cite{Hutschenreuter2019} maps, respectively, may indeed trace the Orion spur. We note that our determination of the pitch angle of the Orion spur may be affected by an angular offset as discussed in the following section.

\subsection{Is there a characteristic spiral arm signature?}
\label{sect:SpiralArmProfile}
The RT simulated $RM$ profile in Fig. \ref{fig:ProfileRM} shows distinct peaks at spiral arm tangents identified in the simulation. The peaks of tangents C, G, and F are distinctly asymmetric, with a much steeper $RM$ gradient on the outside. The $RM$ peaks associated with the local arm (tangents D and E) are broad and symmetric. The $RM$ profiles of the tangent points in the inner galaxy appear blended. In Fig. \ref{fig:ProfileRMComp} we zoom in on the peak associated with tangent point F, which is most similar to the Sagittarius arm tangent observed by \citet{Shanahan2019} in terms of viewing angle with respect to the Galactic centre (and therefore distance from the Sun and the Galactic centre). 

{ The red curve in Fig. \ref{fig:ProfileRMComp} shows the synthetic $RM$ as a function of longitude in relation to the THOR $RM$ data of the region. For the benefit of this paper, $RMs$ from the THOR survey that were derived since the work of \citet{Shanahan2019}, but following their procedure, are also shown as green dots. An extensive THOR $RM$ catalog will be published in a future paper. Also plotted is the synthetic dust profile (blue) for comparison. The synthetic data sets are mirrored  so that they are centered on $l = 48.6^{\circ}$. }

Coming from the Galactic outskirts ($l>48.6^{\circ}$) our RT model reproduces the low $RM$ on the outside of the arm, the peak, and the higher $RM$ for lines of sight that intersect the arm at lower longitude. The peak itself is not as sharp as the THOR data. A similar trend can be seen in the trend for the $870\ \mu\mathrm{m}$ dust profile.

{ 
The THOR $RM$s indicate a pattern of enhanced mean $RM$ and enhanced $RM$ variance for lines of sight through the arm ($l\lesssim 46^{\circ}$) compared with lines of sight that do not intersect the arm (l $\gtrsim 50^{\circ}$), as noted by \citet{Shanahan2019}. Near the spiral arm tangent, a strong excess in mean $RM$ and $RM$ variance is seen. The increased $RM$ variance may be the result of small-scale structure such as HII regions in the spiral arm. Note that the synthetic $RM$ profile is expected to reproduce the mean $RM$, not the excesses due to small scale structure. The highest $RM$s up to $4219\ \mathrm{rad\ m}^{-2}$ may be affected by the proximity of the large HII region complex W51 as reported by \cite{Shanahan2019}. The simulations, as represented by the red curve, trace the gradual increase in $RM$ toward lower longitude well. They also show an excess $RM$ at the arm tangent that is significant compared with the $RM$ of the remainder of the disk (see also Figure~\ref{fig:ProfileRM} for wider context). The excess Faraday rotation in the simulations is, however, smaller than indicated by the data.}

To quantify this, we fitted the parameters to the longitudinal fit function\footnote{The parameters as well as the fit function itself were determined with the help of \textsc{Mathematica} Version 11.0, https://www.wolfram.com/mathematica} ${ f(l)=a + b \times \exp\left[ - (c + l)^2 \right] + d \times l^5 }$  to the complete THOR data set (both \citealt{Shanahan2019} data and newly calculated $RM$s). { Indeed, the trend may also be represented by another function e.g. by a polynomial. The point is, that the fit reveals a comparable trend with that of our simulated $RM$ profile  with a steady increase resulting in a maximum and followed by a further increase towards the galactic center. We note that a similar pattern is also present in the dust profile.}

\begin{figure*}
\centering
\begin{minipage}[c]{1.0\linewidth}
     \includegraphics[width=1.0\textwidth]{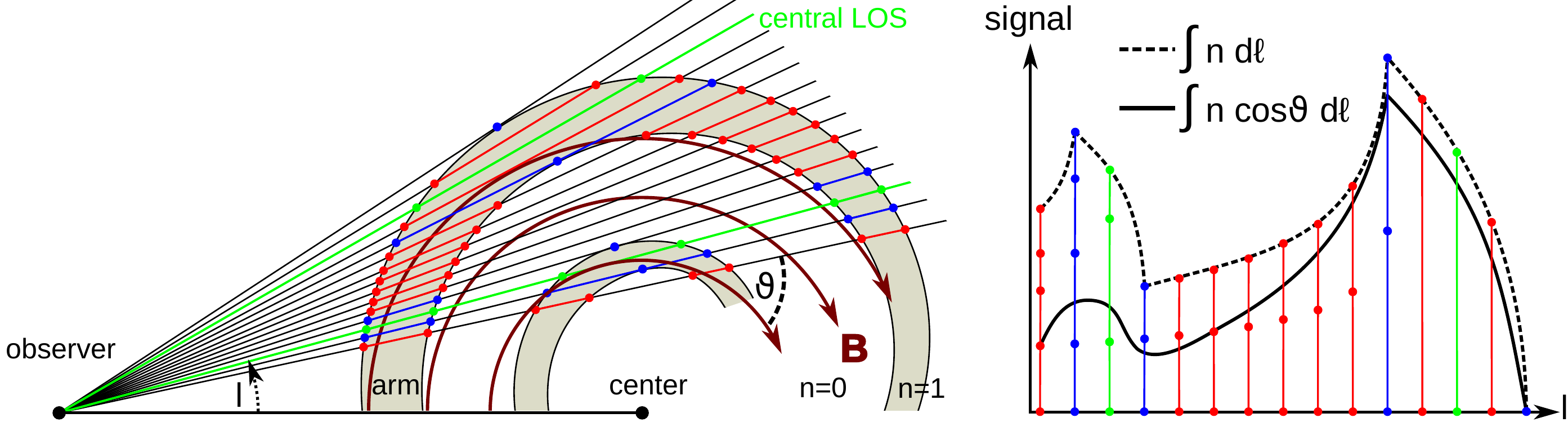}
\end{minipage}

\caption{Schematic sketch of the toy model (left) illustrating the origin of the characteristic shark-fin spiral arm profile (right). Black lines are LOSs in the Galactic plane converging towards the observer, blue lines mark the inner and outer ``tangents'' of the spiral arm, and the green line is the LOS through the center of the arm. The arms have an arbitrary density of $n=1\ \mathrm{a.u.}$ where as the inter-arm regions remain empty. The arms follow a logarithmic spiral where as the direction of the magnetic field $\vec{B}$ is purely toroidal. The distinct LOSs and $\vec{B}$ define the angle $\vartheta$. Coloured dots indicate the intersection of the different LOSs and the arm borders. The diagram on the right shows a re-constructed signal along the longitude l. The dashed black line is calculated assuming that the signal strength is proportional to the length of the portion of the LOSs which lies within the spiral arm. The dashed line is weighted by $\cos \vartheta$. Note that the peak in the observed signal corresponds to the \emph{inner} tangent (blue), and not to the central LOS (green) as usually assumed.}
\label{fig:SketchProj}
\end{figure*}
We speculate that such a shape is primarily a result of the geometry of the arm itself. In order to test this hypothesis we construct a simple 2D arm model consisting of two identical logarithmic spirals. A schematic illustration of this model is provided in Fig. \ref{fig:SketchProj}. The spiral arms are assumed to have a finite width. We assume a constant density of unity within the spiral arms while the inter-arm regions remain empty. Then, we trace rays through the arms towards a the observer position for different equidistant longitudes l. Hence, the tracer signal in this toy model directly scales proportionally to the length of the path which lies within the arm.  

Plotting this signal as a function of the longitude $l$ we find in Fig. \ref{fig:SketchProj} a characteristic shape of a rise and subsequent decline resembling a shark fin. We refer to this as the {\it shark-fin profile}. It has a peak and a shoulder in the direction of the Galactic centre, similar to the profiles of our Galactic model and the data of \cite{Shanahan2019} as well as the dust peaks in Fig. \ref{fig:ProfileSyncDust}. Naturally, the profile of our simple toy model is not exactly what we see in Fig. \ref{fig:ProfileSyncDust} and Fig. \ref{fig:ProfileRMComp} since we ignored secondary effects from the magnetic field, gas density, electron fractions, and dust distribution. Repeating the calculation of the signal weighted by $\cos \theta$ leads to an decrease in magnitude and the second peak corresponding to the inner tangent of the arm becomes smoothed. This peak resembles no longer the {\it shark-fin profile} but is more comparable to those seen in \cite{Oppermann2012} and \cite{Hutschenreuter2019}, respectively. 

In case of the $RM$, the angle of the magnetic field with the line of sight modifies the shark-fin profile somewhat. The main feature to recognize is that for lines of sight that intersect the arm twice, the tangent point where $\vartheta \approx 0$ is absent. The largest effect is therefore in the wing of the shark-fin profile. This is illustrated in Fig. \ref{fig:SketchProj} for a magnetic pitch angle of $0^\circ$.

An interesting aspect that follows from our 2D toy model is that the peak in the profile is not associated with line of sight through the center of the arm but represents the inner border of a spiral arm \citep[see also][]{HouHan2015}. Consequently, all observations of dust, $RM$, and also synchrotron tracers may systematically underestimate the Galactic longitude of the tangent of the spine of a spiral arm by a few degrees (assuming that the quantity of interest is the centre of the spiral arm). 

It may even be possible to derive an arm width from the shape of the profile by similar geometric considerations. However, quantifying the exact parameters for an offset angle or an arm width would require an extensive systematic evaluation of RT models and observational data alike and is therefore beyond the scope of this paper.

\begin{figure*}
\centering
\begin{minipage}[c]{1.0\linewidth}
     \includegraphics[width=0.51\textwidth]{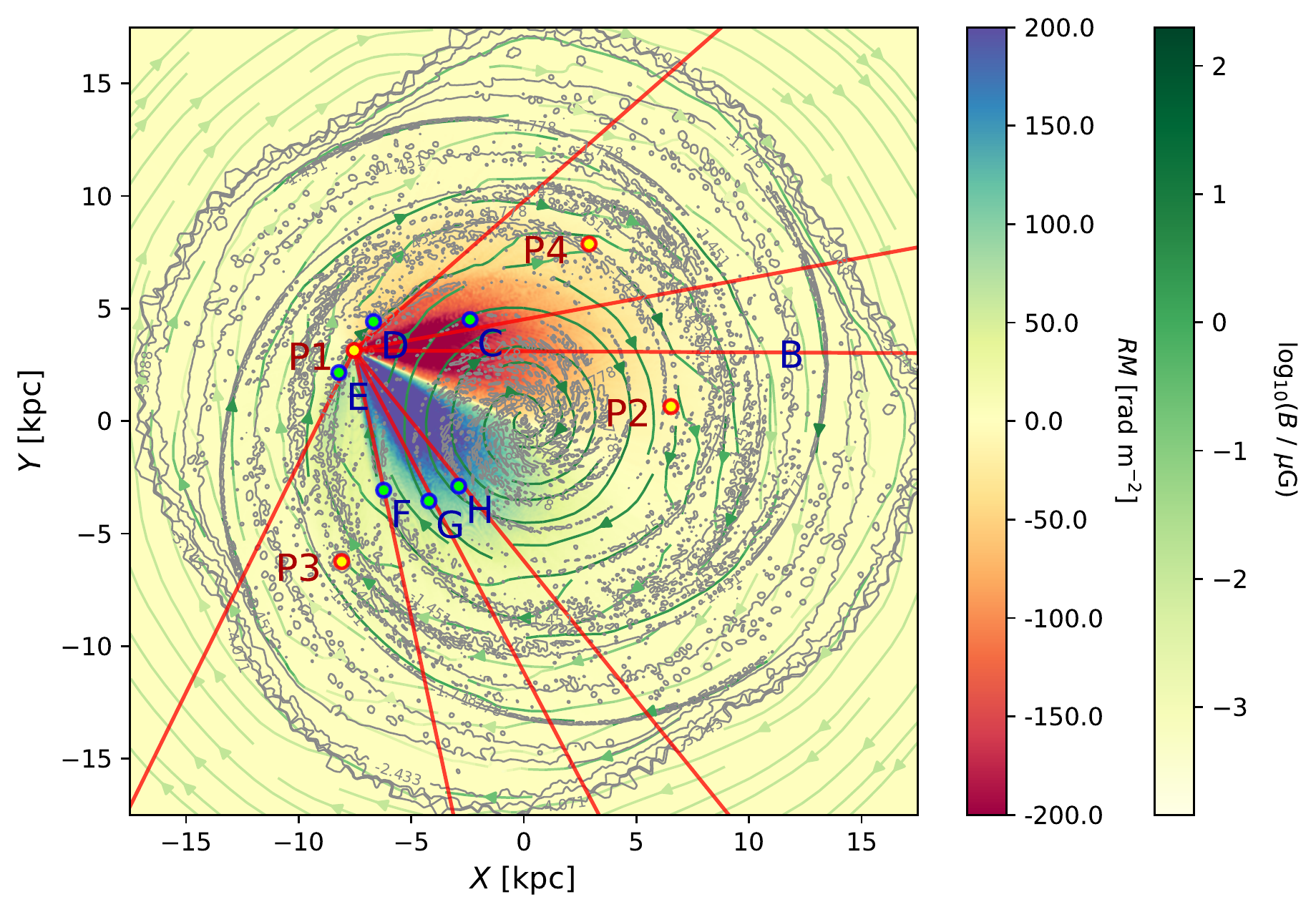}     \includegraphics[width=0.51\textwidth]{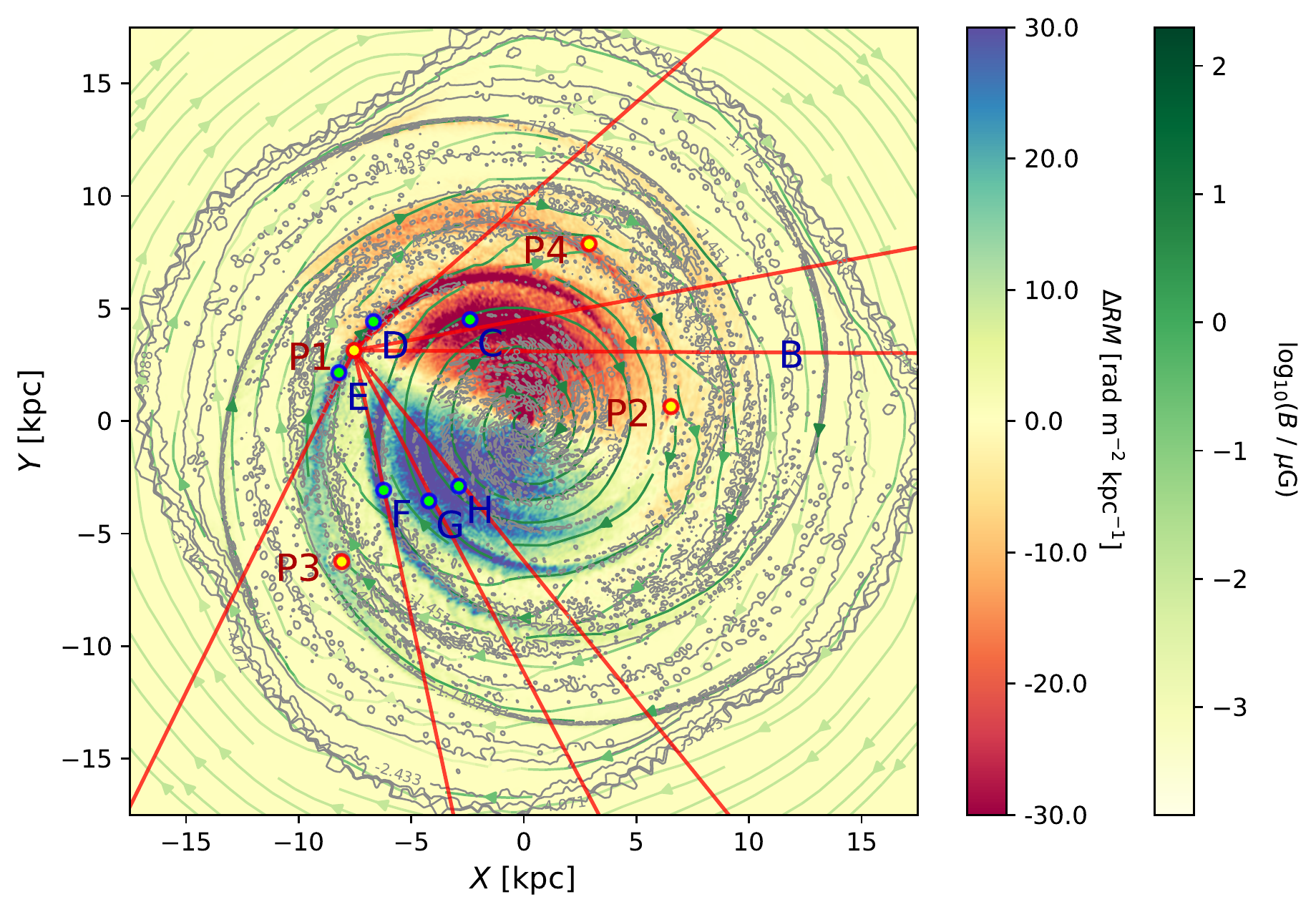}
\end{minipage}
\caption{Left panel: Accumulated values of the $RM$ along all LOSs that lie within the Galactic disk plane. All LOSs start outside the MHD grid with a $RM=0\ \mathrm{rad\ m}^{-2}$ and incrementally converge with each step size $\mathrm{d}\ell$ inwards at the very observer position P1 where the $RM$ reaches its maximum. The magnitude of the resulting $RM$ signal at position P1 is shown in Fig. \ref{fig:ProfileRM} in the left panel. Gray contours indicate the electron density and the vector field represents the magnetic field strength and orientation (see Fig. \ref{fig:Midplanes}). Red lines are the tangents B - H detected by the $RM$ signal where as blue dots represent the tangent points derived from the column density map shown in Fig. \ref{fig:NHMaps}. Right panel: The same as the left panel. Here, we plot the relative change ${\Delta RM = \left(RM(\ell + \mathrm{d}\ell)-RM(\ell)\right)/\mathrm{d}\ell }$ along all LOSs as they approach P1.}
\label{fig:AccRM}
\end{figure*}

\subsection{The origin of the $RM$ signal}
\label{sect:OriginRM}

We repeat the RT post-processing for the $RM$ runs. In the previous runs all rays start outside the grid and converge at the very positions of the observers. Now, we stop the RT runs at the surface of a sphere surrounding the observer with a distinct pre-defined radius. This way, the signal represents the accumulated contribution of the $RM$ up to this particular sphere. Hence, incrementally increasing the radius of the sphere allows us to investigate the origin of the $RM$ signal within our Milky Way model.

In Fig. \ref{fig:AccRM} we present the accumulated increase of the $RM$ signal towards the observer P1 as well as the relative change $\Delta RM$ between adjacent spheres in the $XY$-midplane of the Galactic disk simulation. The accumulated signal coming from the direction of the Galactic center starts to reach a considerable amount of $RM$ within a distance of roughly $10\ \mathrm{kpc}$ away from P1. Hence, we note that Fig. \ref{fig:AccRM} has to be taken with care since other $RM$ modelling efforts as  presented e.g. in \cite{Pakmor2018} indicate that a significant part of the $RM$ signal may even originate at about $15\ \mathrm{kpc}$ away from the observer. However, in our model the accumulation of $RM$  is mostly due to the increase in the magnetic field strength near the center of our Milky Way model (see also Fig. \ref{fig:Midplanes}). Consistent with the $\cos\vartheta$-dependency of the Faraday $RM$ (see Eq. \ref{eq:RM}) we see no signal at all from the very center of the galaxy.

Faraday $RM$ is significantly enhanced in the spiral arms when compared with inter-arms regions. This becomes even more obvious in the plot of the relative change $\Delta RM$ of the signal shown in Fig. \ref{fig:AccRM}. The distribution of $\Delta RM$ within the galactic disk matches the spiral structure of the Milky Way model. We also note some asymmetry for the observer position P1 along the tangents D and E, respectively, with a smaller increase $\Delta RM$ in direction of tangent E. This is also an effect of the magnetic field strength since the strength increases as the arm spirals from the outskirts towards the center. This result is also consistent with the longitudinal profile of $RM$ in Fig. \ref{fig:ProfileRM} where the peak of the tangent D is larger than that of E. In conclusion, the peaks C-H of the longitudinal $RM$ signal are  indeed associated with the tangent points within the spiral arms.

The $RM$ signal is linearly dependent on electron density $n_{\mathrm{el}}$ and the LOS magnetic field strength $B_{\mathrm{||}}$. However, we find a Pearson correlation of $r= 0.91$ between $|\Delta RM$| and $|B_{\mathrm{||}}|$ but only a $r= 0.10$ between $|\Delta RM|$ and $n_{\mathrm{el}}$ for the disk midplane shown in Fig. \ref{fig:AccRM}. Hence, the spiral arm detection by means of $RM$ observations in our Milky Way model is mostly a consequence of the  increase of magnetic field strength within the arms. We emphasise that this particular aspect of our MHD model may not accurately represent the Milky Way where the density of free electrons $n_{\mathrm{el}}$ is also found to be higher within the arms \citep[see e.g.][]{Cordes2002,Langer2017}.  Thus, the $RM$ signal in the Milky Way may be much more influenced by electron fractions than in our synthetic observations suggest. 

\section{The impact of the Local Bubble}
\label{sect:LocalBubble}

In this section we discuss the influence of the local conditions of the observers on our synthetic observations. The sensitivity of synthetic Galactic synchrotron and line observations to the presence of such a Local Bubble has also been discussed in \cite{Reissl2019} and \cite{Pellegrini2019}. These authors found that if an observer is not surrounded by a bubble, i.e.\ a low density cavity, then nearby molecular clouds may completely dominate the tracer signal seen by that observer over large patches of the sky. This situation might prevent any detection of spiral arms. 

As outlined in Sect. \ref{sect:SimSetup} we forced the creation of bubbles into the spiral arms at distinct positions by adjusting the local SN parameters of the MHD simulation. The emerging bubbles are approximately $8\ \mathrm{kpc}$ away from the Galactic center and roughly ${ 50-300\ \mathrm{pc} }$ wide in order to provide conditions comparable to the Local Bubble surrounding our own Solar neighbourhood \citep[][]{Fuchs2009,Liu2017,Alves2018}. We place the observer at the four distinct positions labeled P1, P2, P3, and P4 (see Fig. \ref{fig:Midplanes}). Each position represents a different stage of the bubble evolution. 

Position P1 is surrounded by a cavity with a central gas density of $n_{\mathrm{g}}\approx 0.1\ \mathrm{cm}^{-3}$, an extension of roughly ${ 100-150\ \mathrm{pc} }$, and is the most similar to our Local Bubble \citep[][]{Fuchs2009,Liu2017,Alves2018}. P1 is in a semi-static stage between the expansion and collapse phase. Consequently, the velocity field and the magnetic field lines are well ordered. The resulting profiles have peaks above the noise level and the CO velocity has well defined loops (see Figs. \ref{fig:VelMap} and \ref{fig:ProfileRM}), which allow the detection of tangent points. 

\begin{figure*}
\centering
\begin{minipage}[c]{1.0\linewidth}
     \includegraphics[width=0.478\textwidth]{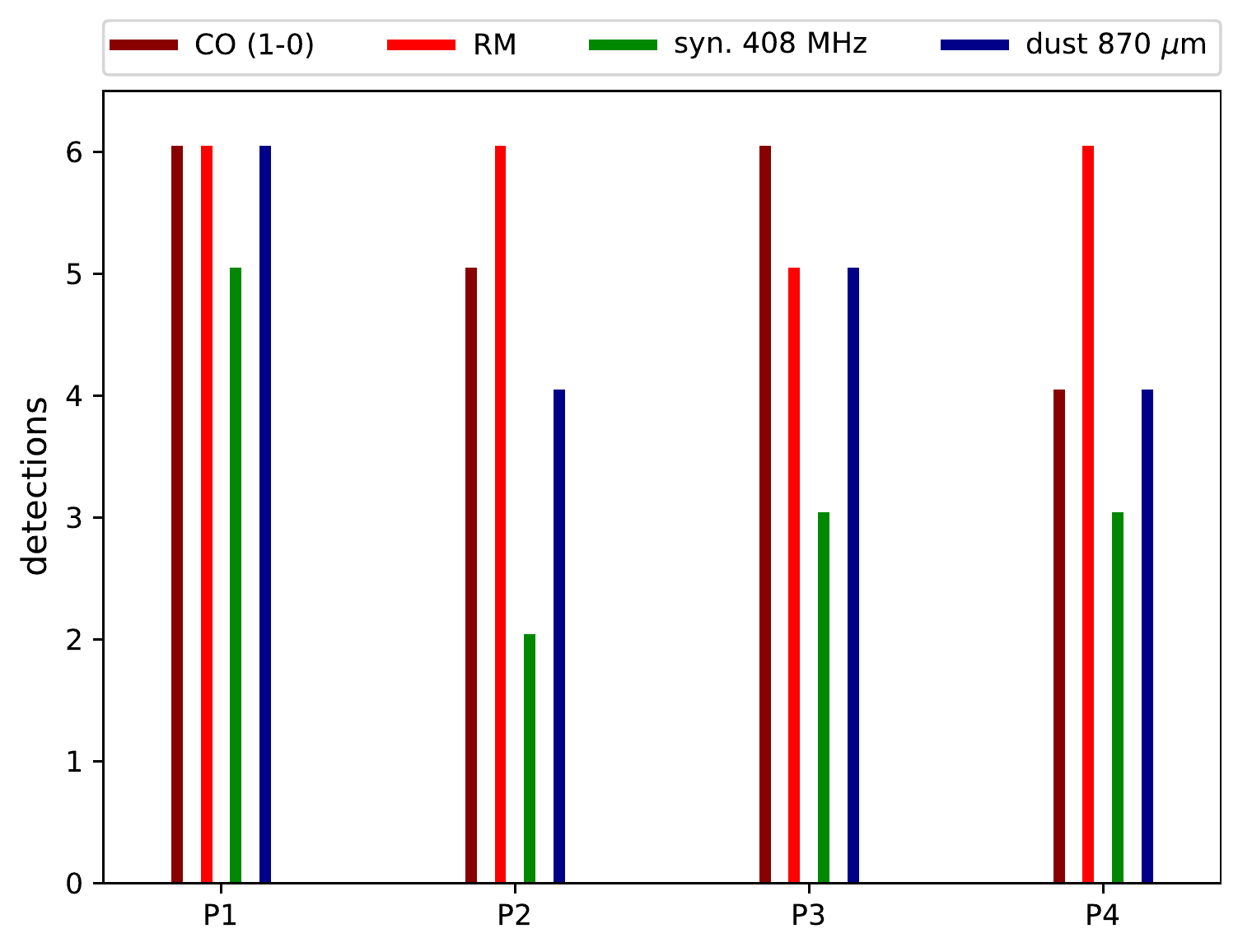}     \includegraphics[width=0.49\textwidth]{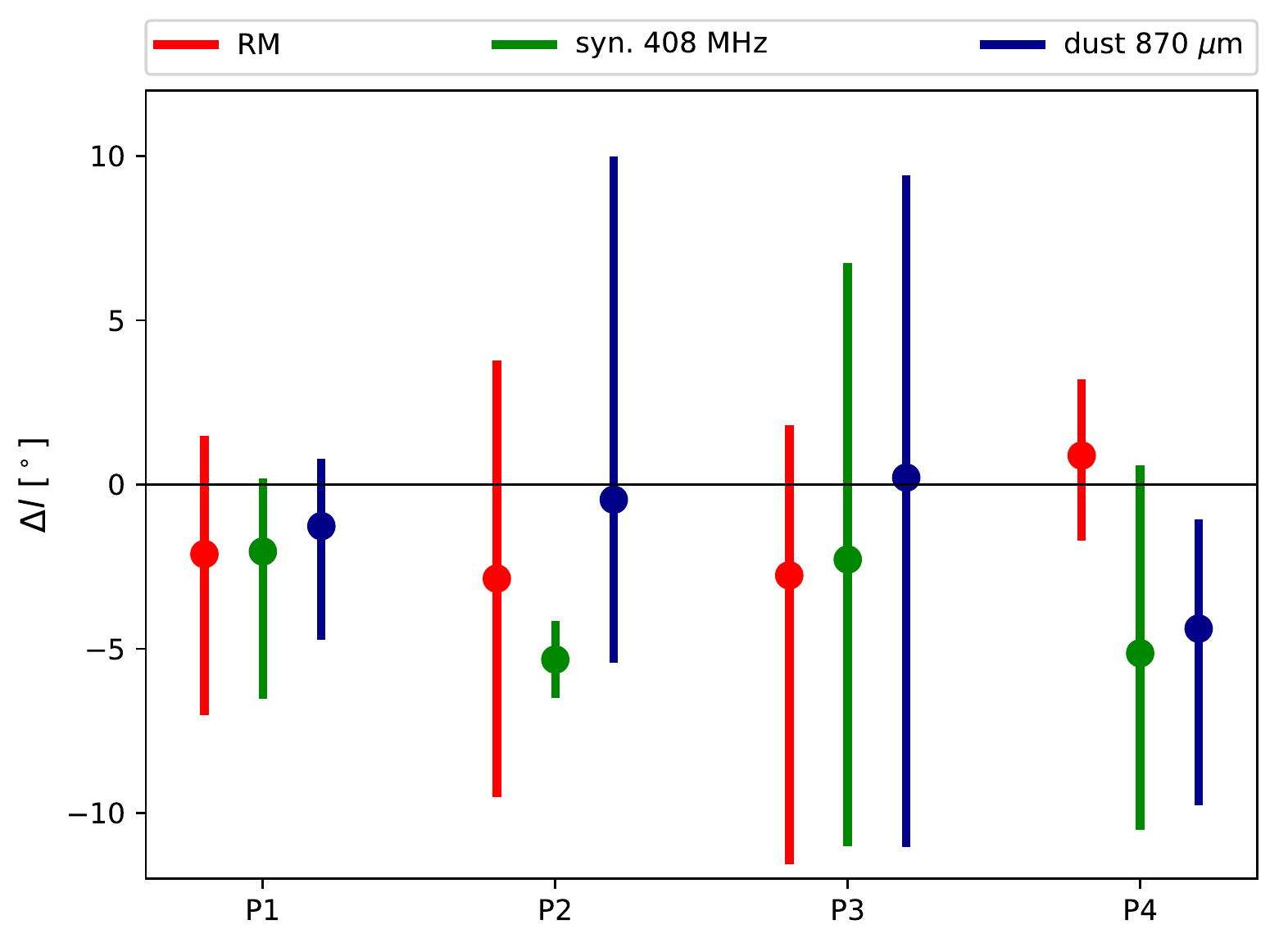}
\end{minipage}

\caption{Left: Number of spiral arm detections per tracer for the different observer positions P1, P3, P3, and P4, respectively. Here, only the tangent points C - H are considered leading to a maximum of six possible detections. Right: The deviation of tangent point angles with respect to the angle of CO of all spiral arms. Vertical lines represent the minimal and maximal deviation for each of the four observer positions P1-P4. The colors correspond to the tracers considered in this paper and the dots are the average value.}
\label{fig:ArmDist}
\end{figure*}

In Fig. \ref{fig:ArmDist} we show the number of spiral arm detections for each observer position. Here, we only count the detections of the points C - H that are clearly separated from the blurry disk in the center of the MHD simulation. With the exception of synchrotron, all tracers can recover the tangent points C - H. We calculate the relative deviations for each tracer and tangent angle point with respect to the CO data. We plot the maximal and minimal deviation of all tangent points in Fig. \ref{fig:ArmDist}. For P1, the maximal appearing deviation to the CO angle is $- 6.88^{\circ}$. We report a systematic offset of roughly $-2^{\circ}$ for all tracers. \citet{HouHan2015} found the offsets between gaseous tracers of spiral arm tangencies to be of this magnitude. We speculate that this systematic offset may also be related to the offset for the tracers of synchrotron, dust, and $RM$ discussed in Sect. \ref{sect:SpiralArmProfile}.

Position P2 is already in its collapse phase in which the cavity is filling up with gas. The local density is almost the same as in the local arm of P2. The velocity field and gas density distribution are much more turbulent than at the location of P1. Indeed, the loops of the local arm are broken in the longitude-velocity diagram and one of the local arm tangents cannot be detected. This also affects the synchrotron emission and the dust signal. Due to the increase in noise, some of the tangent points can no longer be detected. The only tracer that allows us to recover all six tangent points is the $RM$. Here, the range of deviations from the CO tangents is much larger compared to P1 with angles up to $9.85^{\circ}$. Only the synchrotron detections seem to agree with CO. However, for P2 the synchrotron detection is limited and shows only detectable signs of two of the six tangent points. {  We emphasize that the low number of detections in the synchrotron signal may be biased by our simple distribution of CR electrons For actual detections of tangents within the Milky Way synchrotron observations may still be a tracers of tangents.}

The bubble embedding P3 has just recently been formed and is still expanding. Hence, the magnetic field is increased compared to the P1 and P2 position and the gas surrounding the bubble is highly turbulent. The bubble itself is almost completely evacuated with a density in the order of $n_{\mathrm{g}}\approx 10^{-4}\ \mathrm{cm}^{-3}$ and a maximal extension of about ${300\ \mathrm{pc} }$. P3 is also not that close to the spine of a spiral arm but placed in a diffuse inter-arm region. Despite these difficulties all tangent points are detectable with CO observations whereas all other tracers are missing some tangent point detection. Here, with values up to $-11.42^\circ$  the deviations to the detected angles by CO are even larger compared to that of P1 and P2, respectively.

Finally, the observer at P4 is placed directly into a spiral arm unaffected by any SN feedback. Hence, the gas density is very smooth with  well behaved magnetic field lines and a regular velocity field. For P4, the longitude-velocity diagram shows a broad luminous band within $\pm 25\ \mathrm{km\ s}^{-1}$ emerging from the local environment. This band prevents the detection of the tangent points of the local arm of P4. The remaining arms C, F, G and H are clearly separated in velocity space. Dust and synchrotron emission are also heavily influenced by features of nearby clouds and characteristic peaks related to any tangent points become buried in noise. Only, the $RM$ allows to detect all tangents. We speculate that this may be due to the smooth electron fraction and magnetic field surrounding position P4. The deviations of the tangent angles with values up to $-10.38^{\circ}$ are generally smaller than those of P2 and P3. 

One may have intuitively expected that an inner-arm location without any bubble may provide the worst  environment for observing its own galaxy. However, as Fig. \ref{fig:ArmDist} reveals, the turbulence of gas and magnetic field fluctuations seem also to be  important factors. In essence, the conditions of P1 represent the most ideal environment for spiral arm detections within the set of observer positions considered in this paper.

\section{Caveats}
\label{sect:caveats}

There are several limitations in our Milky Way simulation that potentially have an impact on the synthetic observations presented in this paper. First, the cosmic ray ionization rate of $\zeta_{\rm H} = 3 \times 10^{-17} \: {\rm s^{-1}}$ adopted in the simulations of \citet{Smith2020} and Chen et al. (in prep.) is somewhat lower than the values indicated by observations of H$_{3}^{+}$, OH$^{+}$, H$_{2}$O$^{+}$ and ArH$^{+}$ in the diffuse ISM \citep{Indriolo2012,Indriolo2015,Neufeld2017}, which are more consistent with a value an order of magnitude higher. Therefore, the simulations will tend to underestimate the electron density in regions where cosmic ray ionization dominates by a factor of $\sqrt{10} \sim 3$. Since ${\rm RM} \propto n_{\rm el}$, this means that we may underestimate the actual $RM$ values by a factor of a few. Note, however, that this will only affect the overall normalization of our $RM$ values and not their relative magnitude as a function of l. Therefore, this uncertainty should have no effect on the structure we see in the $RM$ profiles.

Second, as our simulation does not account for photoionization feedback from massive stars, it will miss any contribution from ionized gas in and on the boundaries of  HII regions to the measured $RM$ values. We should therefore not expect good agreement between our results and observations of $RM$ towards known Galactic HII regions \citep[see e.g.][]{Purcell2015}. However, the high values of $RM$ found by \citet{Shanahan2019} around l$ = 48^\circ$ are largely found away from HII regions and hence it is meaningful to compare these with our simulation results.

 Third, our neglect of radiative feedback from massive stars also means that we will underestimate the dust temperature and hence the 870~$\mu$m continuum flux in the vicinity of young stellar clusters. However, this should not have a large impact on our results. At 870~$\mu$m, we are in the Rayleigh-Jeans tail of the dust emission spectrum and so the brightness of the emission varies only slightly faster than linearly with the dust temperature.  Since the dust temperature varies with the dust heating rate $\Gamma_{\rm dust}$ as $T_{\rm d} \propto \Gamma_{\rm dust}^{\alpha}$ with $\alpha = 1/5$--1/6, depending on the dust properties, even large spatial variations in the dust heating rate introduce only small variations into the dust
 temperature and the 870~$\mu$m flux.

Finally, our model of CO formation and destruction is highly approximate and has been shown by \citet{Glover2012} to overproduce CO in molecular cloud conditions. In addition, even though the spatial resolution of the Chen et al.\ simulations is excellent by the standards of simulations of entire galactic disks\footnote{Compare, for instance, Chen et al.'s resolution of $\sim 1$~pc at $n = 100 \: {\rm cm^{-3}}$ with the values at a similar resolution of 8~pc in \citet{Fujimoto2019} or $\sim 20$~pc in \citet{Kortgen2019}.}, it is not sufficient to yield a completely numerically converged CO distribution \citep{Joshi2019}. We would therefore expect some of the small-scale details of the CO distribution to change if we increased the numerical resolution of the simulation or adopted a more accurate treatment of the CO chemistry. However, the large-scale distribution of CO emission in position-velocity space should remain largely unaffected, and so we would not expect the results presented in this paper to be sensitive to our choice of chemical model or our numerical resolution in dense regions. 

\section{Summary and Conclusion}
\label{sect:Summary}
We present a comparison of synthetic spiral arm detections and observations of the Milky Way within the Galactic plane. Synthetic observations are created based on a MHD simulation of the galactic disk. The simulation provides a Milky Way-like spiral structure and includes gas self-gravity, chemistry, and SN feedback. In a post-processing step we synthesize spiral arm observations by mimicking the RT physics of cold dust emission, synchrotron radiation, line emission, and Faraday rotation. We analyse the synthetic longitudinal profiles for each individual tracer in order to detect the characteristic signatures of spiral arm tangents. 
This physical model for Faraday rotation in the Milky Way is compared with Faraday rotation from the THOR survey to investigate the contribution of spiral arms to the total rotation measure of the Milky Way.

\begin{itemize}
    \item We report an angular separation of about $\pm 6^\circ$ of tangent points observed in different tracers. This separation is similar to that of the Milky Way.
    \item Our models confirm that the pitch angle of the Milky Way's spiral arms may be recovered from its tangent points on the basis of trigonometrical considerations. The recovered range of angles in our models is in agreement with the pitch angles observed in the Milky Way.

    \item The $RM$ signal is traced through the simulations along distinct LOSs as it accumulates on its way towards the observer. Here, we find that spiral arms create characteristic peaks in Galactic Faraday rotation with contrast and longitude extent that depend on viewing geometry and Galactocentric radius of the tangent point. We offer an explanation of such a profile based on a simple geometrical model of a spiral galaxy.
    \item We find characteristic peaks in the $RM$ signal observed in the Milky Way that may be attributed to the local arm (Orion spur) where our solar system is situated. The pitch angle corresponding to such peaks agrees with the angles given in the literature. Hence, this finding provides an additional means to validate the orientation of the Orion spur within the Milky Way.

    \item The pattern of a sharp peak in Faraday rotation flanked by low $RM$ on the outside, and elevated $RM$ on the inside as reported by \citet{Shanahan2019} is reproduced by the simulations. Other spiral arms may show a broader peak depending on the distance of the tangent point from the observer and from the Galactic centre.

    \item We repeat our analysis for observers situated in distinct SN bubbles. Such bubbles represent a density cavity in different stages of their development within the MHD spiral arms.  Our analysis reveals that a Local Bubble most similar to our own solar neighborhood provides the best Galactic environment for the detection of spiral arm tangents.   
\end{itemize}

The analysis of the resulting synthetic observations lays the groundwork for the interpretation of current and future observations of the Milky Way and their comparison with the theory of galaxy evolution.

\begin{acknowledgements}
Special thanks goes to Claire Rajkay for useful discussions. S.R., S.C.O.G., R.S.K., M.R.R. and H.B. acknowledge funding from the Deutsche Forschungsgemeinschaft (DFG, German Research Foundation) -- Project-ID 138713538 -- SFB 881 ``The Milky Way System'' (sub-projects A06, B01, B02, and B08) and from the Priority Program SPP 1573 ``Physics of the Interstellar  Medium''  (grant  numbers  KL  1358/18.1,  KL  1358/19.2). S.R., S.C.O.G. and R.S.K. acknowledge support from the DFG via the Heidelberg Cluster of Excellence {\em STRUCTURES} in the framework of Germany’s Excellence Strategy (grant EXC-2181/1 - 390900948). The authors gratefully acknowledge the data storage service SDS@hd supported by the Ministry of Science, Research and the Arts Baden-Württemberg (MWK) and the German Research Foundation (DFG) through grant INST 35/1314-1 FUGG and support by the state of Baden-Württemberg through bwHPC and the German Research Foundation (DFG) through grant INST 35/1134-1 FUGG.

J.M.S acknowledges the support of the Natural Sciences and Engineering Research Council of Canada (NSERC), 2019-04848. R.J.S. acknowledges an STFC Ernest Rutherford fellowship(grant ST/N00485X/1)and HPC from the Durham DiRAC super-computing facility. H.B. and J.D.S. acknowledge support from the European Research Council under the Horizon 2020 Framework Program via the ERC Consolidator Grant CSF-648505. 
\end{acknowledgements}


\bibliographystyle{aa}
\bibliography{./bibtex}
 \end{document}